\documentclass[fleqn,usenatbib]{mnras}

\usepackage{newtxtext,newtxmath}

\usepackage[T1]{fontenc}

\DeclareRobustCommand{\VAN}[3]{#2}
\let\VANthebibliography\thebibliography
\def\thebibliography{\DeclareRobustCommand{\VAN}[3]{##3}\VANthebibliography}

\usepackage{graphicx}	
\usepackage{amsmath}	
\usepackage{amssymb}	
\usepackage{wasysym}    
\usepackage{mathtools}  
\usepackage{caption}
\usepackage{subcaption}
\usepackage{array,multirow,graphicx}  

\newcommand{\Icrit}{I_{\rm c}}
\newcommand{\teq}{t_{\rm eq}}
\newcommand{\zeq}{z_{\rm eq}}
\newcommand{\angf}{\omega_{\rm 21}}
\newcommand{\ar}{A_{\rm r}}
\newcommand{\br}{B_{\rm r}}

\newcommand{\lya}{Ly\,$\alpha$}

\newcommand{\fstarII}{f_{\rm *,II}}
\newcommand{\fstarIII}{f_{\rm *,III}}
\newcommand{\vc}{V_{\rm c}}
\newcommand{\fX}{f_{\rm X}}
\newcommand{\emin}{E_{\rm min}}
\newcommand{\tdel}{t_{\rm delay}}

\newcommand{\post}{\mathcal{P}}
\newcommand{\prior}{\pi}
\newcommand{\like}{\mathcal{L}}
\newcommand{\Z}{\mathcal{Z}}

\newcommand{\likeh}{\like_{\rm HERA}}
\newcommand{\likes}{\like_{\rm SARAS}}
\newcommand{\likex}{\like_{\rm X-ray}}

\newcommand{\nuis}{\theta_{\rm n}}
\newcommand{\astro}{\theta_{\rm a}}
\newcommand{\foreg}{\theta_{\rm f}}

\newcommand{\erf}{\textrm{erf}}

\title[21-cm Constraints on Cosmic Strings]{On the Constraints on Superconducting Cosmic Strings from \\21-cm Cosmology}

\author[T. Gessey-Jones et al.]{T. Gessey-Jones,$^{1,2}$\thanks{E-mail: tg400@cam.ac.uk}
S. Pochinda,$^{1,2}$
H. T. J. Bevins,$^{1,2}$
A. Fialkov,$^{2,3}$
W. J. Handley,$^{1,2}$
E. de Lera Acedo,$^{1,2}$\newauthor
S. Singh,$^{4}$
and R. Barkana$^{5}$
\\
$^{1}$Astrophysics Group, Cavendish Laboratory, J. J. Thomson Avenue, Cambridge, CB3 0HE, UK\\
$^{2}$Kavli Institute for Cosmology, Madingley Road, Cambridge, CB3 0HA, UK\\
$^{3}$Institute of Astronomy, University of Cambridge, Madingley Road, Cambridge, CB3 0HA, UK\\
$^{4}$Raman Research Institute, C V Raman Avenue, Sadashivanagar, Bangalore 560080, India\\
$^{5}$School of Physics and Astronomy, Tel-Aviv University, Tel-Aviv 69978, Israel 
}

\date{Accepted XXX. Received YYY; in original form ZZZ}

\pubyear{2023}

\begin{document}
\label{firstpage}
\pagerange{\pageref{firstpage}--\pageref{lastpage}}
\maketitle

\begin{abstract}
Constraints on the potential properties of superconducting cosmic strings provide an indirect probe of physics beyond the standard model at energies inaccessible to terrestrial particle colliders.
In this study, we perform the first joint Bayesian analysis to extract constraints on superconducting cosmic strings from current 21-cm signal measurements while accounting rigorously for the uncertainties in foregrounds and high redshift astrophysics. 
We include the latest publicly available 21-cm power spectrum upper limits from HERA, 21-cm global signal data from SARAS~3, and the synergistic probe of the unresolved X-ray background in our final analysis. 
This paper thus constitutes the first attempt to use 21-cm power spectrum data to probe cosmic strings. 
In contrast to previous works, we find no strong constraints can be placed on superconducting cosmic strings from current 21-cm measurements.
This is because of uncertainties in the X-ray emission efficiency of the first galaxies, with X-ray emissivities greater than $3 \times 10^{40}$\,erg\,s$^{-1}$\,M$_{\odot}^{-1}$\,yr able to mask the presence of cosmic strings in the 21-cm signal. 
We conclude by discussing the prospects for future constraints from definitive 21-cm signal measurements and argue that the recently proposed soft photon heating should be cause for optimism due to its potential to break degeneracies that would have otherwise made the signatures of cosmic strings difficult to distinguish from those of astrophysical origin.
\end{abstract}




\section{Introduction}

Many extensions to the standard model predict the existence of cosmic strings.
Should cosmic strings exist they can play a significant role in early structure formation~\citep{Brandenberger_1994}, a topic receiving renewed attention due to the recent launch of JWST and the detection of overly-large galaxies at high redshifts~\citep[e.g.][]{Labbe_2023, Boylan-Kolchin_2023, Akins_2023}. 
In addition, cosmic strings may also explain the excess radio background seen by ARCADE~2 and the LWA~\citep{Fixsen_2011, Dowell_2018, Cyr_2023b}. 
Hence, any probes of these topological defects allow for the indirect constraint of new physics at high energies, and can potentially provide insight into outstanding problems in astrophysics and cosmology.

One promising current and near-future probe is the 21-cm signal from between the cosmic dark ages and reionization. 
The 21-cm signal measures the excess or deficit of rest-frame 21-cm wavelength photons due to the neutral hydrogen gas present throughout the intergalactic medium in these early epochs~\citep{Madau_1997, Furlanetto_2006, Pritchard_2012, Barkana_2016, Mesinger_2019}. 
Through the spatial and time evolution of the 21-cm signal, the formation of the first stars and galaxies is traceable, allowing for insight into early astrophysics~\citep[e.g.][]{Yajima_2015, Cohen_2016, Mirocha_2018, Mebane_2018, Tanaka_2018,  Schauer_2019, Mebane_2020,  Munoz_2022, Gessey-Jones_2022, Bevins_2022} and the nature of dark matter~\citep[e.g.][]{Barkana_2018, Munoz_2018, Fraser_2018, Liu_2019, Jones_2021, HERA_theory_22, Barkana_2023}.

Multiple mechanisms have been proposed by which cosmic strings can impact the 21-cm signal. 
The overdensity produced in the wake of a string enhances the 21-cm signal in a wedge~\citep{Brandenberger_2010, Hernandez_2012, Hernandez_2014} which should be visible in 21-cm images and three-point statistics~\citep{Maibach_2021}. 
Furthermore, a subclass of cosmic strings, those that carry supercurrents~\citep{Witten_1985}, may enhance the 21-cm signal globally through the emission of an excess radio background~\citep{Feng_18, Theriault_2021}. 
This latter effect potentially being of sufficient magnitude to allow for preliminary constraints on superconducting cosmic strings from the disputed EDGES 21-cm signal measurement~\citep{EDGES, Hills_2018, Brandenberger_2019, Singh_2019, Sims_2020}. 

The field of observational 21-cm cosmology has seen rapid development since the EDGES measurement.
With the SARAS~3 global signal null-result~\citep{SARAS3} rejecting the best-fit EDGES profile at the 95.3\% level, and the latest public HERA 21-cm power spectrum upper limits~\citep{HERA_obs_22, HERA_obs_23} which are now sufficiently low to be probing the astrophysics of the first galaxies~\citep{HERA_theory_22}. 
Combining these global 21-cm signal and 21-cm power spectrum results has proven fruitful, giving stronger constraints on a class of 21-cm models with an excess radio background from high redshift radio galaxies~\citep{Bevins_2023}.
These observations and the subsequent analyses show that existing 21-cm signal measurements are already teaching us much about the early Universe.
However, they have also revealed the complex degeneracies that exist between high-redshift astrophysical parameters, foregrounds, and exotic physics, which must be carefully considered before drawing any conclusions from these data sets.

Further insights into the early Universe have come from the unresolved X-ray background~\citep{Brandt_2022}, the residual X-ray flux measured by X-ray telescopes after subtraction of resolved point sources. 
This unresolved background acts as an upper bound on the X-ray flux from high redshift sources, which thus constrains high redshift astrophysical parameters~\citep{Fialkov_2017}.
\citet{HERA_obs_23} and \citet{Pochinda_2023} have shown that such constraints are complementary to those from the 21-cm signal.

In this paper, we build on the work of \citet{Pochinda_2023} (an extension of an earlier analysis by \citet{Bevins_2023}), whose authors showed that a multi-wavelength analysis combining observations of the 21-cm power spectrum~\citep[HERA Phase 1 Limits,][]{HERA_obs_23}, 21-cm global signal~\citep[SARAS~3 null result,][]{SARAS3}, unresolved X-ray background~\citep{Hickox_2006, Harrison_2016}, and excess radio background~\citep{Dowell_2018}, gave significant insights into early Universe astrophysics, including providing constraints on the properties of galaxies and Population III stars.
Instead of focusing on astrophysics, here we investigate the constraints on cosmic strings a joint analysis of the same 21-cm power spectrum, 21-cm global signal, and unresolved X-ray background data can provide.
We do not include excess radio background data as our cosmic string model is not valid at $z = 0$.
This paper thus constitutes the first analysis to attempt to constrain superconducting cosmic strings using 21-cm power spectrum data. 

We first alter the version of 21-cm Semi-numerical Predictions Across Cosmic Epochs~\citep[\textsc{21cmSPACE},][]{Visbal_2012, Fialkov_2012, Fialkov_2013, Fialkov_2014, Fialkov_2014b, Cohen_2016, Fialkov_2019, Reis_2020, Reis_2021, Reis_2022, Magg_2022, Gessey-Jones_2022, Gessey-Jones_2023, Sikder_2023} used in \citet{Pochinda_2023} to model an excess radio background from superconducting cosmic strings.
Using this code, we generate a data set of 21-cm signal and X-ray background predictions for different astrophysical and cosmic strings scenarios, from which we train emulators for efficient evaluation of these observables. 
We then employ a Bayesian methodology to allow for the proper marginalisation of the foregrounds and the uncertain astrophysics of the early Universe when extracting our constraints on cosmic strings.
Ultimately, we determine that astrophysical uncertainties are too great for us to draw any robust conclusions about superconducting cosmic strings from existing 21-cm observations.

During the preparation of this paper, a series of related works~\citep{Acharya_2023, Cyr_2023a, Cyr_2023b} were published.
These studies propose an intriguing new mechanism \textit{soft photon heating} by which superconducting cosmic strings heat the intergalactic medium, an effect not included in our study. 
\citet{Acharya_2023} find the inclusion of this heating suppresses the amplitude of the 21-cm global signal.
As a result, \citet{Cyr_2023a} find there to be no constraints on superconducting cosmic strings at $2\sigma$ significance when treating the EDGES global signal measurement as a limit on the 21-cm global signal magnitude at $z = 18$.
This is in agreement with our conclusions that current 21-cm signal measurements cannot constrain superconducting cosmic strings.
However, we reach this conclusion due to the uncertainty in astrophysical heating while their heating is intrinsic to the cosmic strings.
Together our works illustrate the importance of considering all effects, foreground, astrophysical, and from the strings themselves when attempting to constrain cosmic strings using the 21-cm signal. 
We discuss further how including soft photon heating would impact our conclusions in Section~\ref{sec:conclusions}.

We begin by recapping the theory of the excess radio background from superconducting cosmic strings in Section~\ref{sec:excess_radio_background}, and its impact on the 21-cm signal in Section~\ref{sec:21cm}. 
Then we outline the observations we use for our constraints in Section~\ref{sec:methodology} alongside our data analysis methodology. 
Afterwards, the results from our analysis are presented in Section~\ref{sec:results}. 
Finally, we conclude in Section~\ref{sec:conclusions}, with discussions of our findings and with the prospects for future constraints on superconducting cosmic strings from 21-cm signal measurements.

\section{An Excess Radio Background from Superconducting Cosmic Strings}~\label{sec:excess_radio_background}

Throughout this work, we focus on the impact of superconducting cosmic strings on the 21-cm signal via the excess radio background, for which we use the model developed by \citet{Brandenberger_2019} and \citet{Theriault_2021}. 
Below, we recap the main details of this model. A reader familiar with these works may wish to skip to the next \hyperref[sec:21cm]{section}.

If cosmic strings exist, they will have formed before recombination as the Universe is rapidly cooling post-Big Bang~\citep{Brandenberger_1994}.
As the temperature drops, so does the energy scale of the fields that permeate the Universe. 
When this energy scale crosses critical values, the nature of the ground state (also called vacuum) of one or more of these fields changes abruptly in what is called a phase transition~\citep{Mazumdar_2019}. 
If the new ground states no longer have a symmetry present in the old ground state, the transition is a spontaneous symmetry-breaking one.
The canonical example of such a transition is the Higgs mechanism, wherein the electroweak gauge symmetry is broken as the Higgs field undergoes a phase transition at an energy scale of $160$\,GeV~\citep{Onofrio_2016}.

In a symmetry-breaking phase transition, the new ground states of the field will be a degenerate set. 
Causality dictates that the first small regions of the new ground state to form will thus be in different degenerate ground states~\citep{Kibble_1976, Kibble_80, Kibble_82}.
These regions of new vacuum grow outward at the speed of light and eventually fill the entire Universe. 
However, if the topology of the new degenerate ground state is not simply connected (i.e.\ it has holes), these expanding bubbles of new vacuum cannot smoothly merge into one region all in the same vacuum state. 
As a result, where the bubbles meet topological defects in the field are left behind, regions where the field is stuck in a higher energy state because the ground state's topology prevents it from relaxing.
This process, wherein causality and ground state topology combine to create topological defects in cosmological fields, is the Kibble mechanism~\citep{Kibble_1976}.

The nature of the topological defects produced via the Kibble mechanism depends on the specific topology of the new ground states, with possible types of defect including cosmic strings, domain walls, monopoles and texture~\citep{Brandenberger_1994}.
For example, cosmic strings are produced if the set of new ground states is topologically equivalent to a 1D ring (1-sphere). 
These cosmic strings are nearly 1D regions where the field is stuck in a higher energy state.
As a result, a gauge particle condensate exists along the string, thus giving it gravitational mass. 
A network of such strings is expected to form via the Kibble mechanism if an appropriate symmetry-breaking phase transition occurs, comprising a mixture of infinite strings permeating the Universe and a smaller number of finite string loops~\citep{Vanchurin_2006}.
Since none of the phase transitions in the standard model of particle physics produce cosmic strings, their presence would be a smoking gun sign of physics beyond the standard model, allowing any observations that constrain their potential properties to probe these theories.

If the condensate forming the cosmic strings is charged, which is true for a large subclass of theories, the whole string becomes superconducting~\citep{Witten_1985}, and so carries a theory-dependent current $I$. 
Consequently, superconducting strings lose energy to electromagnetic radiation, in addition to the gravitational radiation losses that all cosmic strings experience. 
Due to having two competing energy loss mechanisms superconducting cosmic strings can be split into three categories based on which loss mechanism dominates: supercritical strings that primarily lose energy to electromagnetic radiation; subcritical strings that primarily lose energy to gravitational waves; and critical strings for which the two energy loss mechanisms are of equal magnitude. The critical current which divides these regimes for fixed $\mu$ is given by\footnote{All equations in this section are given in natural units, $c = k_{\rm B} = \hbar = 1$.}
\begin{equation}~\label{eqn:critical_current}
    \Icrit = \kappa^{-1} \gamma G^{-1/2} (G\mu)^{3/2},
\end{equation}
where $G$ is Newton's constant, $\kappa$ a dimensionless electromagnetic emissivity constant of order 1, and $\gamma$ is another dimensionless constant related to the efficiency of gravitational wave emission~\citep[numerical simulations give $\gamma \sim 100$,][]{Vachaspati_1985}. Supercritical strings have $I > \Icrit$, and subcritical strings have $I < \Icrit$.

We will assume that all cosmic strings in our model have the same current $I$ and tension $\mu$, and hence all are either supercritical, critical or subcritical. That is not however to say we model all cosmic strings as identical. The phase transition which produced the cosmic strings will have left the strings with a range of loop radii $R$~\citep[e.g.][]{Vilenkin_1981, Albrecht_85, Hindmarsh_95}, which in the matter-dominated epoch follows the scaling number density
\begin{equation}~\label{eqn:cosmic_string_number_density}
    n(R, t) = 
    \begin{cases}
        \chi R^{-5/2} \teq^{1/2} t^{-2} & \textrm{for } R_{\rm c}(t) < R < \psi t_{\rm eq}, \\
        R^{-2} t^{-2}  & \textrm{for } R \geq \psi t_{\rm eq},
    \end{cases}
\end{equation}
where $t$ is cosmic time, $\teq$ the cosmic time of matter-radiation equality, $\chi$ and $\psi$ are model-specific dimensionless constants,\footnote{The $\chi$ and $\psi$ constants here were called $\nu$ and $\alpha$ in \citet{Brandenberger_2019} but are renamed here to avoid confusion in subsequent sections with frequency and X-ray spectral index respectively.} and $R_{\rm c}(t)$ the cutoff radius. The cutoff radius is defined as the radius below which cosmic strings are expected to decay within a Hubble time and so this value is dependent on the dominant energy-loss mechanism of the cosmic strings
\begin{equation}~\label{eqn:cirtical_radius}
    R_{\rm c}(t) = 
    \begin{cases}
        \kappa \beta^{-1} I \mu^{-1/2} t & \textrm{for } I > \Icrit, \\
        \kappa \beta^{-1} G \mu t  & \textrm{for } I \leq \Icrit.
    \end{cases}
\end{equation}
Here $\beta$ is another dimensionless model-specific parameter, encoding the shape of the loops.\footnote{$2 \pi$ for circular loops.}

The final piece of information required to calculate the radio background from cosmic strings is their emissivity. \citet{Cai_2012} calculated the power $P$ emitted per angular frequency $\omega$ by superconducting cosmic strings at low frequencies to be
\begin{equation}~\label{eqn:dpdw_scs}
    \frac{dP}{d\omega} = \kappa I^2 R^{1/3} \omega^{-2/3}.
\end{equation}
By combining equations~\eqref{eqn:cosmic_string_number_density}, \eqref{eqn:cirtical_radius}, and \eqref{eqn:dpdw_scs} the radio emissivity of the cosmic string population can be found. 
Radio photons began to be able to freely stream through the Universe at recombination, and so the radio background post-recombination is given by integrating the contribution from cosmic strings back to recombination accounting for the redshifting of photons by the expansion of the Universe. The resulting excess energy density in radio photons below a given angular frequency was found by \citet{Brandenberger_2019} to be
\begin{equation}~\label{eqn:rho_t}
\begin{split}
    \rho(t; \omega) = 18 \Tilde{\kappa}& \chi \beta^{7/6} \omega^{1/3} \teq^{1/2}t^{-13/6}   \\
    & 
    \times \begin{cases}
        \kappa^{-7/6} G^{-7/12} I^{5/6} (G \mu)^{7/12}  & \textrm{for } I > \Icrit, \\
        \gamma^{5/6} \kappa^{-2} G^{-1} (G \mu)^{11/6} & \textrm{for } I = \Icrit, \\
        \gamma^{-7/6} I^2 (G \mu)^{-7/6} & \textrm{for } I < \Icrit,
    \end{cases}
\end{split}
\end{equation}
with $\Tilde{\kappa}$ being $\kappa$ multiplied by an order one constant. 

Here we are treating the cosmic string background as homogeneous.
At the redshifts of interest, this assumption begins to break down at $G \mu \gtrsim 1$ when there is no longer a large number of cosmic string radio bubbles overlapping at every point in space. 
In this work, we do not consider $G \mu > 10^{-6}$, as these tensions have been ruled out via pulsar timing arrays~\citep{Miyamoto_2013}.
As a result, for our applications, the homogeneity assumption is well justified. 

From the above, we can find the superconducting cosmic string produced excess radio temperature at the 21-cm line frequency $\angf$
\begin{equation}~\label{eqn:delta_t_rad}
\begin{split}
    \Delta T_{\rm rad}^{\rm 21} = 6 \pi^2 \Tilde{\kappa}& \chi \beta^{7/6} \angf^{-8/3} \teq^{-5/3}\left(\frac{1 + z}{1 + \zeq}\right)^{13/4}  \\
    & 
    \times \begin{cases}
        \kappa^{-7/6} G^{-7/12} I^{5/6} (G \mu)^{7/12}  & \textrm{for } I > \Icrit, \\
        \gamma^{5/6} \kappa^{-2} G^{-1} (G \mu)^{11/6} & \textrm{for } I = \Icrit, \\
        \gamma^{-7/6} I^2 (G \mu)^{-7/6} & \textrm{for } I < \Icrit,
    \end{cases}
\end{split}
\end{equation}
where we have converted from cosmic time to redshift $z$, with $\zeq$ the redshift of matter-radiation equality. Note our formula differs by a factor of 9 to that of \citet{Theriault_2021} due to the authors erroneously converting between $\rho$ and $\Delta T_{\rm rad}$ using the formula for black-body radiation. The cumulative energy spectrum from cosmic strings in the radio is a power-law with spectral index $1/3$ (e.g.~equation~\eqref{eqn:rho_t}) not black-body (approximately a power-law with spectral index $3$ at low frequencies), correcting for which results in the difference in the numeric factor. 

\textsc{21cmSPACE}, discussed in the next section, can model a spatially-homogeneous excess radio background at the rest-frame 21-cm line of the form~\citep{Fialkov_2019, Reis_2020}\footnote{In some previous works~\citep{Fialkov_2019, Reis_2020} $\ar$ was defined as the magnitude of the excess radio background relative to the CMB at $z = 17.2$, rather than at $z = 0$. Hence, there is a numeric factor difference of $18.2^{\br}$ between the $\ar$ convention in these works and ours. For $\br = 2.25$, this numeric factor is $684$, thus, the limit on $\ar$ required to replicate the depth of the EDGES signal found in \citet{Fialkov_2019} corresponds to $\ar \gtrsim 0.0028$ in our notation.}
\begin{equation}~\label{eqn:erb}
    \Delta T_{\rm rad}^{\rm 21} = T_{\rm cmb, 0}(1+z)\left(\ar [1 + z]^{\br}\right),
\end{equation}
with $T_{\rm cmb, 0}$ being the cosmic microwave background temperature seen today. By comparison to the theoretical prediction for the radio background above, we see they are of the same form if we define the relative magnitude of this excess to be
\begin{equation}
\begin{split}
    \ar = 6 \pi^2 \Tilde{\kappa}& \chi \beta^{7/6} \angf^{-8/3} \teq^{-5/3} T_{\rm cmb,0}^{-1}\left(1 + \zeq\right)^{-13/4}  \\
    & 
    \times \begin{cases}
        \kappa^{-7/6} G^{-7/12} I^{5/6} (G \mu)^{7/12}  & \textrm{for } I > \Icrit, \\
        \gamma^{5/6} \kappa^{-2} G^{-1} (G \mu)^{11/6} & \textrm{for } I = \Icrit, \\
        \gamma^{-7/6} I^2 (G \mu)^{-7/6} & \textrm{for } I < \Icrit,
    \end{cases}
\end{split}
\end{equation}
and the exponent of its redshift evolution as
\begin{equation}
    \br = 9/4.
\end{equation}
This $\ar$ formalism has the added advantage of condensing all of the degenerate model-specific parameters, $\beta$, $\gamma$, $\kappa$, $\Tilde{\kappa}$, $\chi$, $\mu$, and $I$ into one. Thus, we can, and will, constrain the phenomenological radio background strength $\ar$ and later convert these constraints to any superconducting cosmic string model of interest. 

A homogeneous radio background also described by equation~\eqref{eqn:erb} has been previously constrained using 21-cm signal data in \citet{HERA_theory_22} and \citet{Bevins_2022}, using an earlier set of HERA Phase 1 21-cm power spectrum upper limits, and the SARAS~3 21-cm global signal data respectively.
With both finding higher excess radio background magnitudes were disfavored but not ruled out by their respective data sets. 
However, in those studies, the background was assumed to be from synchrotron radiation and a value of $\br = 2.6$ was used,\footnote{The value of 2.6 used in \citet{HERA_theory_22} and \citet{Bevins_2022}, was motivated by the measurement of the spectral index of the present-day excess radio background by \citet{Dowell_2018}, $2.58 \pm 0.05$. In the same work, the spectral index of galactic synchrotron radiation was found to vary from $2.48$ to $2.62$ across the sky. So, while there is some uncertainty, or intrinsic variation, in the spectral index of synchrotron radiation its value is larger than the $\br$ value predicted for superconducting cosmic strings.} rather than the value of $2.25$ that we consider here.
Hence, we would expect the radio backgrounds from cosmic strings to have a smoother evolution than that used in those studies, decaying away less rapidly.
As a result, the two types of radio backgrounds will have distinct impacts on the time evolution of the 21-cm signal, thus requiring us to perform a separate analysis rather than reinterpreting the results from these earlier studies. 

We should re-emphasise here for clarity that the above equations assume a matter-dominated epoch. 
This is a good approximation for the period of the Universe's history that the 21-cm signal is sensitive to ($5 < z < 150$).
However, the approximation breaks down at lower redshifts and no longer holds at $z = 0$ due to the effects of dark energy. 
Thus, we cannot use the derived equations to rigorously constrain superconducting cosmic strings using the present-day excess radio background measured by ARCADE~2 and the LWA~\citep{Fixsen_2011, Dowell_2018}. 
If we were to assume equation~\eqref{eqn:delta_t_rad} holds down to $z = 0$, then the excess radio background measurements listed in \citet{Dowell_2018} rule out $\ar \gtrsim 0.1$.
Since we do not anticipate significant deviations from equation~\eqref{eqn:delta_t_rad} for the complete $\Lambda$CDM calculation this figure can be treated as illustrative of the order of magnitude of constraints on $\ar$ from ARCADE~2 and LWA.
But, as this figure is only approximate, we do not use it as part of our constraints or consider the present-day radio background further in this work. 
While this paper was in preparation, \citet{Cyr_2023a} published a thorough analysis constraining the parameters of superconducting cosmic strings using present-day radio background measurements, including a full modelling of $\Lambda$CDM cosmology and a more detailed radiative transfer treatment than is used here.

\section{Impacts of an Excess Radio Background on the 21-cm Signal}~\label{sec:21cm}

The 21-cm signal is the change in abundance of rest-frame 21\,cm wavelength photons relative to a radio background caused by the presence of neutral hydrogen in the early Universe~\citep{Furlanetto_2006}. 
Normally, this signal is expressed as the difference in radio brightness temperature observed today $T_{\rm 21}$.
Due to the expansion of the Universe, the differential brightness temperature at different frequencies probes different epochs of the history of the Universe.
The strength of the signal observed depends on the optical depth of the 21-cm line $\tau_{\rm 21}$, the temperature of the radio background $T_{\rm rad}$ and the spin temperature of the observed patch of hydrogen gas $T_{\rm s}$ as 
\begin{equation}
T_{\rm 21}  = \left(1 - e^{-\tau_{\rm 21}} \right)\frac{T_{\rm s} - T_{\rm rad}}{1+ z}.
\label{eqn:Tb_equation}
\end{equation}
$T_{\rm s}$ is a statistical temperature encompassing the relative occupation of excited and non-excited states of neutral hydrogen. 
Thus the relative values of $T_{\rm s}$ and $T_{\rm rad}$ determine whether the hydrogen gas is a net emitter or absorber of 21-cm photons, and hence the sign of $T_{\rm 21}$.

From equation~\eqref{eqn:Tb_equation} it can immediately be seen that enhancing the radio background $T_{\rm rad}$ would have an impact on the 21-cm signal~\citep{Feng_18, Ewall_18, Fialkov_2019,  Reis_2020}. 
However, $T_{\rm s}$ also depends on $T_{\rm rad}$.
$T_{\rm s}$ is determined by the competing influences of absorption/emission of radiation from the background, collisions of neutral hydrogen, and the Wouthuysen-Field effect~\citep{Field_1958, Wouthuysen_1952}.
The first of which forces $T_{\rm s}$ toward $T_{\rm rad}$, while the other two force it toward the kinetic temperature of the neutral hydrogen gas. 
Hence, a full picture of how the 21-cm signal depends on $T_{\rm rad}$ needs to account for the relative strength of these three processes, which will vary in both time and space.

\begin{figure*}
    \centering
    \includegraphics[width=\textwidth]{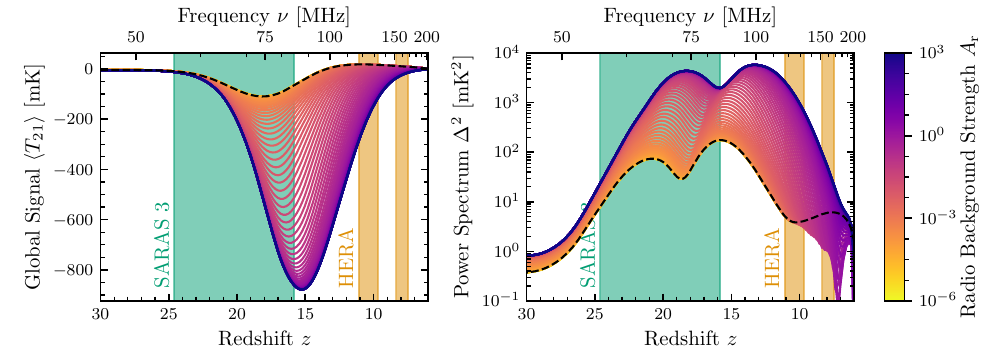}
    \caption{Variation of the 21-cm global signal (left) and 21-cm power spectrum (right) with the strength of the excess radio background. The 21-cm power spectrum is shown for $k = 0.34$\,h\,cMpc$^{-1}$ ($0.23$\,cMpc$^{-1}$). All astrophysical parameters are the same between the depicted signals, and the 21-cm signal when there is no excess radio background (e.g. CMB only) is shown as a black dashed line. The SARAS 3 band is illustrated in green and the HERA bands are in orange. An increase in the excess radio background is seen to increase the magnitude of the global signal absorption trough ($100$\,mK to $900$\,mK) and move it later ($z = 18$ to $ 15$), in this case outside of the SARAS 3 band. Similar effects occur in the 21-cm power spectrum, with the power spectrum magnitude increasing with $\ar$ and the cosmic dawn and heating peaks shifting to later times. At very high  $\ar \gtrsim 10^1$ and very low $\ar \lesssim 10^{-4}$ values of the radio background strength the signal is seen to saturate and no further variation is observed. In the case of very low $\ar \lesssim 10^{-4}$ the signals become visually indistinguishable from the 21-cm signal with no excess radio background. }
    \label{fig:arad_variation_21cm}
\end{figure*}

To accurately calculate the dependence of the 21-cm signal on $\ar$ we utilize our pre-existing simulation code \textsc{21cmSPACE}.\footnote{For a detailed up-to-date description of the code and its latest features see \citet{Gessey-Jones_2023}.}
The code follows a semi-numerical paradigm, with small-scale physics (e.g.\ star formation) handled by analytic prescriptions, and large-scale physics (e.g.\ radiative transfer) modelled numerically.
This hybrid approach speeds up the code considerably compared to numerical simulations, allowing each simulation to complete in a few hours. 
Within the simulation itself, the Universe is modelled as a periodic box, divided into 128$^3$ cells, each with side length 3\,comoving Megaparsecs (cMpc).
As a result, the code ultimately produces 3D maps of the 21-cm signal at this resolution across the range of redshifts the simulation is run between, in our case $z = 50$ to $6$.

However, current 21-cm signal experiments do not aim to produce 21-cm tomographic maps but instead, attempt to observe summary statistics of the 21-cm signal that are easier to measure. 
They target either the sky-averaged 21-cm global signal $\langle T_{\rm 21} \rangle (z)$~\citep{EDGES, PRIZM,  SARAS3, REACH, MIST} or the 21-cm power spectrum $\Delta^2 (z, k)$~\citep{MWA, LOFAR, NenuFAR, HERA_obs_22}.
Consequently, we similarly compress the outputs of all our simulations into these summary statistics.
Both these summary statistics depend on the redshift being observed $z$ and so probe the time evolution of the 21-cm signal.
Additionally, the 21-cm power spectrum probes the spatial evolution of the 21-cm signal through the fluctuations at different scales set by the wavenumber $k$.

Much about the early Universe remains uncertain, and so, \textsc{21cmSPACE} has various parameters and settings that describe or enable astrophysical and cosmological processes.
For all simulations used in this study, we include modelling of:
\begin{itemize}
    \item Baryon dark-matter relative velocities~\citep{Visbal_2012, Fialkov_2012}
    \item Wouthuysen-Field effect~\citep{Fialkov_2013}
    \item \lya{} multiple scattering~\citep{Reis_2021}
    \item \lya{} heating~\citep{Reis_2021}
    \item CMB heating~\citep{Venumadhav_2018, Fialkov_2019}
    \item X-ray heating including SED dependance~\citep{Fialkov_2014b, Pacucci_2014}
    \item Reionization~\citep{Fialkov_2014b}
    \item Photoheating feedback~\citep{Cohen_2016}
    \item Lyman-Werner feedback~\citep{Fialkov_2013, Munoz_2022}
    \item Population III (Pop~III) star to population II (Pop~II) transition~\citep{Magg_2022}
    \item Pop~III star initial mass function~\citep[IMF,][]{Gessey-Jones_2022}
    \item Star formation efficiency suppression in low-mass halos~\citep{Fialkov_2013}
\end{itemize}
We fix the maximum root mean free path of ionizing photons to 40\,cMpc and the Pop III star initial mass function to a  logarithmically flat  IMF between $2$\,M$_{\odot}$ and $180$\,M$_{\odot}$~\citep{Klessen_2023}, as they are found to have little impact on our results.
We assume galaxies do not contribute to the radio background so that there is only one source of excess radio background in our simulations.
Constraining models with multiple sources of excess radio background is left to future works.
The remaining free parameters of the code allow us to explore different scenarios for the uncertain astrophysical properties of the early Universe. 
Specifically, we can vary the excess radio background strength $\ar$ (see previous section), star formation efficiency of Pop~II, $\fstarII,$ and Pop~III, $\fstarIII$, stars, the delay time between the two stellar populations, $\tdel$, the X-ray emission efficiency of galaxies, $\fX$, the X-ray spectrum of galaxies described by a spectral index $\alpha$ and low-energy cutoff $\emin$, the critical circular velocity for stars to form in a halo, $\vc$, and the efficiency of galaxies at emitting ionizing photons, $\zeta$.

We utilize the same version of \textsc{21cmSPACE} as \citet{Pochinda_2023}, but we model a superconducting cosmic string radio background in our simulations rather than an excess radio background from galaxies.
A detailed discussion of the changes to  \textsc{21cmSPACE} from the version commonly employed in older analyses~\citep{HERA_theory_22, Bevins_2022b, Bevins_2023, HERA_obs_23} is provided in that paper.
However, for clarity, we here summarise the relevant improvements to the code.
Star formation is now modelled following the prescription introduced by \citet{Magg_2022}, separating Pop~II and Pop~III star formation.
Hence, halos are now assumed to first form Pop~III stars with efficiency $\fstarIII$ once they cross the critical mass threshold for star formation.
Then the halo takes a time $\tdel$ to recover from the ejection of material by Pop~III star supernovae, after which it can begin forming Pop~II stars with efficiency $\fstarII$.
Each of the stellar populations has a separate Lyman band spectra, in the Pop~III case, this is derived from the Pop~III IMF~\citep{Gessey-Jones_2022}, while for Pop~II stars a fixed spectrum from \citet{Starburst_1999} is used.\footnote{Currently, Pop~II and Pop~III star-forming halos are assumed to have the same X-ray emission efficiency $\fX$ and SED in our models.
This assumption is not anticipated to strongly impact our constraints on $\ar$, and hence on cosmic strings.}
Finally, a module that calculates the X-ray background from high redshift sources has been added to \textsc{21cmSPACE}, following the methodology from \citet{Fialkov_2017}.

For self-consistency, this new X-ray background module calculates the contribution of high-redshift sources to the present-day unresolved X-ray background in a similar manner to which X-ray heating is modelled within \textsc{21cmSPACE}~\citep{Fialkov_2014}.
Hence, it also assumes that star-forming halos are X-ray sources with a starburst galaxy-like luminosity to star formation rate relation~\citep[e.g.][]{Grimm_2003, Mineo_2012}
\begin{equation}
    L_{\rm X} = \left(3 \times 10^{40}\textrm{\,erg\,s}^{-1}\textrm{\,M}_{\odot}^{-1}\textrm{\,yr}\right) f_{\rm X} {\rm SFR},
\end{equation}
where SFR is the star formation rate of the halo and $f_{\rm X} = 1$ corresponds to the specific luminosity predicted by \citet{Fragos_2013} for low metallicity X-ray binaries.
The SED of these X-ray sources is assumed to be a power-law with lower cutoff energy $\emin$ and power-law slope $\alpha$.
By combining the above with the simulated star formation rate density, \textsc{21cmSPACE} is then able to compute the specific X-ray emissivity $\epsilon_{\rm X}(z, E)$ throughout the simulation. 
The present-day specific intensity from $z \geq 6$ sources can then be calculated in the simulation using~\citep[e.g.][]{Pritchard_2007}
\begin{equation}
    J_{\rm X}(E) = \frac{c}{4 \pi} \int_{z' = 6}^{\infty} \epsilon_{\rm X} \left(z', E (1 + z') \right)  \frac{e^{- \tau_{\rm X}(z', E)}}{(1 + z') H(z')} dz',
\end{equation}
where $H$ is the Hubble parameter, $E$ the X-ray energy, and $\tau_{\rm X}(z', E)$ is the optical depth of X-rays between their emission redshift of $z'$ and the present day. 
$\tau_{\rm X}(z', E)$ is, in turn, calculated by integrating the X-ray cross-section of hydrogen and helium species\footnote{The contribution of metals is assumed to be small due to their lower abundance.} in the IGM between $z'$ and $z = 0$~\citep{Verner_1996}.
$J_{\rm X}(E)$ is now a standard output of \textsc{21cmSPACE} simulations to facilitate comparisons, like those we perform in this paper, of models of the early Universe to the observed unresolved X-ray background data.

By fixing the astrophysical parameters and running simulations varying the strength of the excess radio background $\ar$ we can gain an intuition for the impacts of the presence of superconducting cosmic string on the 21-cm signal. 
We illustrate 21-cm signals with $\ar$ varying between  $10^{-6}$ and $10^{3}$ for fixed $\fstarII = 0.05$, $\fstarIII = 0.002$, $\tdel = 30$\,Myr, $\fX = 1$, $\alpha = 1.5$, $\emin = 200$\,eV, $\vc = 5.6$\,km\,s$^{-1}$, and $\zeta = 15$ in Fig.~\ref{fig:arad_variation_21cm}.
Concentrating first on the global 21-cm signal, we observe that as $\ar$ increases, the 21-cm global signal absorption trough becomes deeper, ranging from $100$\,mK to $900$\,mK. 
Depending on the astrophysical parameters chosen, the signal depth under an excess radio background can be up to several Kelvin (for an example, see Fig.~\ref{fig:global_functional_posterior}).
However, this growth is not endless since the signal magnitude eventually saturates~\citep{Reis_2020}; in the illustrated case, this occurs at $\ar \gtrsim 10^1$. 
The fundamental cause of this saturation is the balance between the couplings that determine $T_{\rm s}$.
At very high $\ar$ the radiative coupling dominates, causing $T_{\rm s}$ to become approximately proportional to $T_{\rm rad}$, and since $\tau_{\rm 21} \propto T_{\rm s}^{-1}$ then in this limit $\tau_{\rm 21} \propto T_{\rm rad}^{-1}$.
Hence, the $T_{\rm rad}$ dependence of the bracketed term in equation~\ref{eqn:Tb_equation} and the numerator cancel in the high $\ar$ and $T_{\rm rad}$ limit, leading to a 21-cm signal that no longer depends on $T_{\rm rad}$ or $\ar$~\citep[see][for the exact limit]{Fialkov_2019}.
Conversely, if the excess radio background is much smaller than the CMB, then it has negligible impact on the 21-cm signal as is seen for $\ar \lesssim 10^{-4}$, and so the 21-cm signal is insensitive to $\ar$ in the low $\ar$ limit as well.
One final impact of increasing $\ar$ is that the global signal minimum shifts to later times (lower redshifts), this is potentially important to our study as it can move the global signal minima outside the SARAS~3 band (see subsection~\ref{ssec:saras_3}), as shown in the figure, making the signal harder to distinguish from the smooth galactic foregrounds.

We observe similar trends in the $k = 0.34$\,h\,cMpc$^{-1}$ 21-cm power spectrum.\footnote{h is the Hubble parameter normalized by 100\,km\,s$^{-1}$\,Mpc$^{-1}$, for which throughout we assume the \textit{Planck}~2018~\citep{Planck_VI} best-fit value for h of 0.674. The stated wavenumber is thus equivalent to $0.23$\,cMpc$^{-1}$.}
Like the magnitude of the global signal absorption trough, the magnitude of the 21-cm power spectrum grows with $\ar$ but saturates at both low and high $\ar$ values.
In addition, as $\ar$ increases the cosmic dawn and X-ray heating peaks in the power spectrum move to later times. 
Hence an excess radio background from superconducting cosmic strings can have considerable impacts on the 21-cm signal, suggesting it may be possible to extract constraints on superconducting cosmic strings from existing 21-cm signal data.
We now describe said 21-cm signal data as well as that from complementary probes of the early Universe that we will use in our joint analysis.

\section{Data sets and Methodology}~\label{sec:methodology}

In previous studies, \citet{Bevins_2023} demonstrated that a joint analysis of 21-cm global signal and power spectrum data was able to constrain the properties of the first galaxies, while \citet{HERA_obs_23} and \citet{Pochinda_2023} showed that high-redshift astrophysical constraints could be further improved by including information on the unresolved X-ray background.
Motivated by these works we use a joint analysis of 21-cm global signal, 21-cm power spectrum, and unresolved X-ray background data for our constraints on superconducting cosmic strings, all three of which can be modelled self-consistently by \textsc{21cmSPACE}.
We begin this section by introducing these three sets of observations in subsections~\ref{ssec:hera}, \ref{ssec:saras_3}, and \ref{ssec:xray_background}, before detailing our data analysis methodology in subsection~\ref{ssec:analysis_methodology}.

\subsection{HERA}~\label{ssec:hera}

The Hydrogen Epoch of Reionization Array (HERA) is a radio interferometer designed to detect the 21-cm power spectrum~\citep{HERA_obs_22}. 
It is currently in operation, and the collaboration behind the experiment recently published upper limits on the 21-cm power spectrum based on Phase 1 of their observations. 
Their best current published limits, from 94 nights of observations~\citep{HERA_obs_23}, were $\Delta^2 \leq 457$\,mK$^2$ at $k = 0.34$\,h\,cMpc$^{-1}$ and $z = 7.9$, and $\Delta^2 \leq 3496$\,mK$^2$ at $k = 0.36$\,h\,cMpc$^{-1}$ and $z = 10.4$.
An earlier set of limits published by HERA from 18 nights of observations~\citep{HERA_obs_22} were at the time the most-constraining published power spectrum limits on permitted early Universe scenarios~\citep{Bevins_2023}, able to rule out a range of astrophysical and new physics scenarios.
Hence, we expect the latest limits, which we use in this study, to similarly provide moderate constraints on our astrophysical parameters, and potentially $\ar$.

As HERA employs a foreground avoidance strategy, these limits are directly on the 21-cm power spectrum, and so no foreground modelling is needed to extract constraints from them.
However, there is the potential for residual systematics in the limits leading to excess power above thermal noise. 
We account for such systematics in the same manner as \citet{HERA_obs_23}  through the form of our likelihood, discussed further in subsection~\ref{ssec:analysis_methodology}. 
We are thus inheriting the assumption that residual systematics can only add power, which may not be the case~\citep[e.g.][]{Kolopanis_2019}. 
Known sources of signal loss were corrected for in the computation of the HERA upper limits.

\subsection{SARAS~3}~\label{ssec:saras_3}
The Shaped Antenna measurement of the background RAdio Spectrum (SARAS) experiments~\citep{saras2, SARAS3} are a series of radiometers targeting a detection of the global 21-cm signal.
SARAS~3 is the most recent in the series, a monopole antenna deployed on a lake in Southern India for 14 days and operating in the 43.75 to 87.5\,MHz frequency range.
The experiment was the first of its type to be deployed on a body of water, with this change being made to boost the antenna's overall efficiency and provide a well-characterized uniform medium under the antenna.

We use 15 hours of data from the SARAS3 antenna.
This data has been pre-processed to calibrate the receiver and antenna bandpass, correct for thermal emission from the water below the antenna, and remove radio frequency interference. 
The resulting science data set covers 55 to 85\,MHz (redshift 15.7 to 24.8) in 470 frequency bins, and should consist of a combination of time-averaged foregrounds, the global signal and residual noise.
Previous studies have shown the SARAS~3 foregrounds are well fit by a 7-coefficient polynomial, with a joint fit of a 21-cm signal model and this foreground model able to refute the best-fit EDGES measurement at the 95.3\% level~\citep{SARAS3} and rule out a large swathe of models with enhanced radio backgrounds from galaxies or synchrotron emission~\citep{Bevins_2022}.
We hence adopt a similar approach of jointly fitting our 21-cm global signal predictions alongside a 7-coefficient polynomial foreground model to the data.

\subsection{Cosmic X-ray background}~\label{ssec:xray_background}

The cosmic X-ray background is the unresolved X-ray flux detected by X-ray astronomy experiments~\citep[see][for a recent overview]{Brandt_2022}, calculated by subtracting all known sources from the total flux measured by these experiments.  
As observation times increased and the angular resolution of X-ray telescopes improved, more X-ray point sources have been resolved~\citep{Hickox_2006}. 
Consequently, the level of the unresolved background has decreased over time. 
The remaining unaccounted-for flux should be from a combination of still unresolved point sources~\citep{Harrison_2016} and diffuse emission. 

Except for the brightest quasars~\citep[e.g.][]{Medvedev_2020}, X-ray sources from the epoch of reionization, such as X-ray binaries, are too small and faint to be resolved. 
As a result, the redshifted emission of these sources that survives until today will form a part of the unresolved cosmic X-ray background. 
Hence, measurements of the X-ray background provide upper limits on the X-ray emissivity of these sources, and thus, another way to constrain high redshift astrophysics~\citep{Fialkov_2017}. 

\citet{HERA_obs_23} and \citet{Pochinda_2023} find cosmic X-ray background constraints complement 21-cm signal constraints well, ruling out different regions of the astrophysical parameter space.
So, we also use unresolved X-ray background measurements~\citep{Harrison_2016} from the \textit{Chandra}~\citep{Hickox_2006}, \textit{HEAO-1}~\citep{Marshall_1980, Gruber_1999}, \textit{BeppoSAX}~\citep{Frontera_2007}, \textit{INTEGRAL}~\citep{Churazov_2007} and \textit{SWIFT}~\citep{Ajello_2008} satellites, in our joint constraints.
We list the values used in Table.~\ref{tab:ixrb_bands}.

\begin{table}
 \caption{Collated measurements of the integrated unresolved cosmic X-ray background from \citet{Hickox_2006} and \citet{Harrison_2016}. These measurements are of the total X-ray flux seen by experiments minus the contribution from known sources. Hence, they act as upper limits on the X-ray background from high redshift sources and so can be used to constrain our models of early Universe astrophysics.}
 \label{tab:ixrb_bands}
 \centering
 \begin{tabular}{cc}
  \hline
  Band & Measurement\\
  $\left[\textrm{keV}\right]$ & $\left[\textrm{erg\,cm}^{-2}\textrm{\,s}^{-1}\textrm{\,deg}^{-2}\right]$ \\
  \hline
  1 $-$ 2 & $(1.04 \pm 0.14) \times 10^{-12}$   \\ 
  2 $-$ 8 & $(3.4 \pm 1.7) \times 10^{-12}$   \\  
  8 $-$ 24 & $(1.832 \pm 0.042) \times 10^{-11}$   \\ 
  20 $-$ 50 & $(2.0 \pm 0.083) \times 10^{-11}$   \\ 
  \hline
 \end{tabular}
\end{table}

\subsection{Bayesian analysis methodology}~\label{ssec:analysis_methodology}

For our analysis, we follow a Bayesian methodology similar to that of \citet{Bevins_2023} and \citet{Pochinda_2023}.
In said methodology our \textit{a priori} belief in the probability of different parameter values being true $P(\theta)$, is updated by the likelihood of observing some data $D$ given said parameters $P(D | \theta)$, to give us the \textit{a posteriori} probability of different parameter values being true $P(\theta | D)$, via Bayes' theorem
\begin{equation}~\label{eqn:bayes}
    P(\theta | D) = \frac{P(D | \theta) P(\theta)}{P(D)}.
\end{equation}
Where $P(D)$ is the Bayesian evidence
\begin{equation}
    P(D) = \int P(D | \theta) P(\theta) d \theta.
\end{equation}
As is typical in Bayesian analyses for brevity we denote the quantities in equation~\eqref{eqn:bayes} as the prior $\pi(\theta) = P(\theta)$, the likelihood $\like(\theta) = P(D | \theta)$, the posterior $\post(\theta) = P(\theta | D)$, and the evidence $\Z = P(D)$.

In this study we are principally interested in the constraints we can extract on superconducting cosmic strings, and hence on $\ar$.
The parameters describing uncertain high redshift astrophysics ($\astro$) or foregrounds ($\foreg$) are considered nuisance parameters in this work.
One advantage of adopting a Bayesian approach is we can marginalize over such nuisance parameters to recover the posterior on $\ar$ alone
\begin{equation}~\label{eqn:marg_post}
    \post(\ar) = \frac{1}{\Z} \int \like(\ar, \astro, \foreg) \prior(\ar, \astro, \foreg) d \astro d\foreg,
\end{equation}
where we have expanded the previously general $\theta$ into three classes of parameters. 
In practice, to evaluate $\post(\ar)$ we use the slice-sampling-based nested sampling implementation \textsc{Polychord}~\citep{Handley_2015a, Handley_2015b}.
As a bonus, this approach also allows us to extract astrophysical constraints discussed in more detail in appendix~\ref{app:astro}.
To fully define equation~\eqref{eqn:marg_post} we hence need to specify $\like(\ar, \astro, \foreg)$ and $\prior(\ar, \astro, \foreg)$. 
Let us begin with $\like$.

We adopt the same form for the likelihood of HERA 21-cm power spectrum observations as was used in \citet{HERA_theory_22}
\begin{equation}~\label{eqn_like_hera}
    \likeh(\ar, \astro) \propto \prod_{i}^{n} \frac{1}{2} \left(1+\erf\left[\frac{d_{\rm i} - m_{\rm i} (\ar, \astro)}{\sqrt{2(\sigma_{{\rm d}, i}^2 + \sigma_{{\rm m}, i}^2)}}\right] \right).
\end{equation}
Above $\erf$ is the error function, $n$ the number of data points, $m_{\rm i}$ the model prediction for the power spectrum at the relevant redshift and wavenumber, $\sigma_{{\rm d}, i}$ the standard deviation of the measured power spectrum $d_{\rm i}$, and $\sigma_{{\rm m}, i}$ an optional theory model term. 
This form is derived from assuming that each independent data point measured $d_{\rm i}$ is composed of the true signal plus an unknown systematic $u_{\rm i}$, then marginalizing away the systematic dependence over a uniform prior.  
Since the neighbouring wavenumber bins of HERA overlap, to ensure this assumption of independent data points is valid we only include every other wavenumber bin in our analysis starting with the lowest.

As integrated X-ray background observations act as upper limits on the X-ray background from high redshift sources we follow the methodology of \citet{Pochinda_2023} and adopt the same form as equation~\eqref{eqn_like_hera} for the X-ray background likelihood $\likex$.
The independent data points are now the integrated X-ray background in the four bands outlined in Table.~\ref{tab:ixrb_bands}.

The SARAS~3 data includes a foreground from galactic radio emission which is several orders of magnitude larger than the 21-cm signal. 
These foregrounds are anticipated to be smooth and so should be well-modelled by a low-order log-log polynomial. 
As was done in \citet{SARAS3} and \citet{Bevins_2022}, we model the foreground radio temperature $T_{\rm fg}$ as a 7-coefficient log-log polynomial of frequency $\nu$ 
\begin{equation}~\label{eqn:foregrounds}
    \log_{\rm 10}\left[ \frac{T_{\rm fg}(\nu)}{1{\rm\,K}} \right] = \sum_{i = 0}^{i=6} a_{i} \left(f_{\mathcal{N}}\left[\log_{\rm 10} \left(\frac{\nu}{1{\rm\,MHz}}\right)\right] \right)^i,
\end{equation}
where $f_{\mathcal{N}}$ is a linear normalization to map the $\log_{\rm 10}(\nu / 1{\rm\,MHz})$ values between -1 and 1, and the $a_{i}$ parameters are the foreground coefficient that we treat as nuisance parameters, referred to as a collective as $\foreg$. 
We thus model the sky-averaged radio temperature as a combination of $T_{\rm fg}(\nu; \foreg)$ and the global 21-cm signal $T_{\rm 21}(\nu; \ar, \astro)$ predicted by simulations. 
The noise on the SARAS~3 data is assumed to be Gaussian~\citep{SARAS3} with a standard deviation $\sigma$, which we will also fit as nuisance parameters. 
Hence, the resulting form for the SARAS~3 likelihood we reach is
\begin{equation}~\label{eqn:like_saras}
\begin{split}
    \likes(\ar&, \astro, \foreg) = \prod_{j}^{l} \frac{1}{\sqrt{2 \pi \left(\sigma^2 + \sigma_{\rm th}^2 \right)}} \exp \bigg(\\
    &-\frac{1}{2} \bigg[ \frac{T_{\rm obs, j} - T_{\rm fg}(\nu_j; \nuis ) - T_{\rm 21}(\nu_j; \ar, \nuis)}{\sigma^2 + \sigma_{\rm th}^2} \bigg]^2 \bigg).
\end{split}
\end{equation}
Here we are taking the product over the $l$ SARAS~3 observed frequencies $\nu_{\rm j}$, with corresponding observed sky temperature $T_{\rm obs, j}$. 
We also include an optional theory error term $\sigma_{\rm th}$ to take into account modelling uncertainties.

Constraints for individual experiments are thus found by using the appropriate likelihood for $\like$ in equation~\eqref{eqn:marg_post}. 
Since the data sets considered are statistically independent we can also simply perform joint analyses by taking $\like$ to be the product of the likelihoods we want to include. 
So for our full joint analysis,
\begin{equation}
\begin{split}
    \like(\ar, \astro, \foreg) &= \likeh(\ar,  \astro) \times \\
     &\likes(\ar, \astro, \foreg) \times \likex(\ar,  \astro).
\end{split}
\end{equation}

An individual nested sampling run can require millions of likelihood evaluations,\footnote{For example, our joint constraints run performed 130 million likelihood evaluations.} each of which would na\"ively call \textsc{21cmSPACE}. 
As the code takes a couple of hours to execute this would be prohibitively slow. 
To avoid this problem, we use neural network emulators of \textsc{21cmSPACE} to reduce the evaluation time of the likelihood to tens of milliseconds, making our nested sampling runs possible in a reasonable time frame.
Since we are considering a model with a superconducting cosmic string radio background, which is different to the models used in previous studies, we need to train new emulators for our application. 
We hence ran 44,836 simulations of the early Universe with randomly sampled $\ar$ and astrophysical parameters, the outputs of which are split to form the training and testing set for the emulators of each of our three observables.

We employ the \textsc{Tensorflow}~\citep{tensorflow} based \textsc{globalemu}~\citep{global_emu} as the basis for our global signal emulator, using five dense hidden layers of size 32, early-stopping to avoid overfitting, and a 1 to 10 test-train split. 
The training data used spanned $z = 6$ to $28$, covering both the SARAS~3 and HERA bands.
Once trained the emulator had an average root-mean-square error of 25\,mK for the test data set, which is sufficient for our purposes as the global 21-cm signals we anticipate the SARAS~3  data may be able to rule out are of order 1000\,mK~\citep{Bevins_2023}.

Our power spectrum and X-ray background emulators are derived from the power spectrum emulator developed for \citet{HERA_theory_22} and described in appendix~B thereof. 
They are implemented using the multi-layer perceptron regression neural network from \textsc{scikit-learn}~\citep{sklearn}, adopting a \textsc{globalemu} style methodology of taking as inputs to the network the redshift $z$ and wavenumber $k$ to evaluate the power spectrum at, or energy $E$ to evaluate the X-ray background at. 
The training and testing data sets covered  $z = 6$ to $28$ and $k = 0.085$ to $1.000$\,cMpc$^{-1}$ for the power spectrum emulator, and $E = 0.40$ to $55.00$\,keV for the X-ray background emulator.
Both networks have four hidden layers of size 100 and a 1 to 10 test-train split. 
The converged power spectrum emulator had a relative accuracy of 17\% at the 95\% percentile, comparable to~\citet{HERA_theory_22}, whereas the X-ray background emulator had a 95\% percentile relative accuracy of 5\%.

For the inputs to our emulators, instead of using the phenomenological ionizing efficiency of galaxies $\zeta$ which is input into \textsc{21cmSPACE}, we used the derived optical depth to reionization $\tau$ output by the simulations.
This change was made due to the easier interpretability on $\tau$ and existing constraints on $\tau$ from \textit{Planck}~\citep{Planck_VI} allowing for more reasonable priors on its value.

To account for the aforementioned small emulator errors in our analysis we include a fractional theory/emulator error term $ \sigma_{{\rm m}, i} = 0.17 m_{\rm i}$ in $\likeh$, and $ \sigma_{{\rm m}, i} = 0.05 m_{\rm i}$ in $\likex$. While for the global-signal emulator, we use a frequency-independent theory/emulator error term of $\sigma_{\rm th} = 25$\,mK. This conservative approach should ensure our results are robust to any imprecisions introduced due to using an emulator rather than \textsc{21cmSPACE} directly.

\begin{table}
 \caption{Parameter priors used in our analysis. $\ar$ and the astrophysical parameters are used in all our constraints, whereas the SARAS parameters are only included when the SARAS~3 data is a part of the constraint.}
 \label{tab:prior_table}
 \centering
 \begin{tabular}{ccccc}
  \hline
  Type & Parameter & Prior & Minimum & Maximum \\
  \hline
  $\ar$ & $\ar$ & Log-Uniform & $10^{-6}$ & $10^{3}$ \\
  \hline 
   \parbox[t]{2mm}{\multirow{8}{*}{\rotatebox[origin=c]{90}{Astrophysical}}} & $\fstarII$ & Log-Uniform & $10^{-3}$ & $0.5$ \\
   & $\fstarIII$ & Log-Uniform & $10^{-3}$ & $0.5$  \\
   & $\tdel$ & Log-Uniform & 10\,Myr & 100\,Myr \\ 
   & $\fX$ & Log-Uniform & $10^{-3}$ & $10^{3}$ \\
   & $\alpha$ & Uniform & $1$ & $1.5$ \\
   & $\emin$ & Log-Uniform & 0.1\,keV & 3\,keV \\
   & $\vc$ & Log-Uniform & 4.2\,km\,s$^{-1}$ & 100\,km\,s$^{-1}$  \\
   & $\tau$ & Uniform & 0.033 & 0.075\\
  \hline
   \parbox[t]{2mm}{\multirow{8}{*}{\rotatebox[origin=c]{90}{SARAS}}} & $\sigma$ & Log-Uniform & $0.01$\,K & $1$\,K  \\
   & $a_{\rm 0}$ & Uniform & 3.54 & 3.55 \\
   & $a_{\rm 1}$ & Uniform & -0.23 & -0.21  \\
   & $a_{\rm 2}$ & Uniform & 0 & 0.01  \\
   & $a_{\rm 3}$ & Uniform & -0.01 & 0  \\
   & $a_{\rm 4}$ & Uniform & 0 & 0.01  \\
   & $a_{\rm 5}$ & Uniform & -0.01 & 0.01 \\
   & $a_{\rm 6}$ & Uniform & -0.01 & 0.01 \\
   \hline
 \end{tabular}
\end{table}

Finally, we need to define $\pi(\ar, \astro, \foreg)$ to fully specify our problem. 
We place independent priors on each of our parameters as detailed in Table~\ref{tab:prior_table}. 
In all of our constraints, we include the $\ar$ and {\it Astrophysical} parameters, whereas the {\it SARAS} parameters are only needed when the SARAS~3 data forms part of the constraint. 
The $\tau$ prior is derived from the \textit{Planck} 2018 3-$\sigma$ constraints~\citep{Planck_VI}. 
For the SARAS~3 parameters, we have used priors zoomed in around the best-fit values of \citet{Bevins_2023}. 
We expect the values of the foreground parameters to change little between their work and ours, since \citet{Bevins_2023} found that their astrophysical parameters and foreground parameters are uncorrelated, as well as due to the large-scale difference between foreground and signal. 
As we are only interested in the shape of the posteriors and not the Bayesian evidence this zooming in does not affect our final results and acts to reduce the Kullback–Leibler divergence between prior and posterior reducing the nested sampling runtime~\citep{Ashton_2022}. 
Note, that we treat $\alpha$, $\emin$ and $\tdel$ as continuous parameters like \citet{Bevins_2022} rather than discrete parameters or fixing their values as was done in previous works~\citep{Bevins_2022b, HERA_obs_23, Pochinda_2023}, allowing the emulator to interpolate between the discrete values provided by the simulation data set.

With all terms in equation~\eqref{eqn:marg_post} fully specified we can now present the constraints we find from individual experiments and our joint analysis.

\section{Results}~\label{sec:results}

In this section, we present the parameter constraints from individual experiments, before advancing to our joint constraints. 
We then discuss our marginalized posterior on $\ar$ and the interpretation thereof.
Finally, we will consider our $\ar$ constraints in the context of a superconducting cosmic string model.

\begin{figure*}
    \centering
    \begin{subfigure}[t]{0.49\textwidth}
	\includegraphics{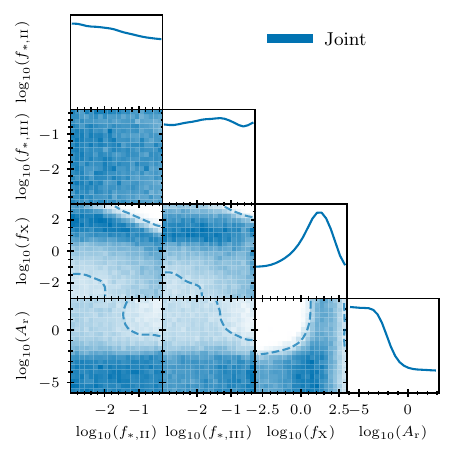}
        \phantomsubcaption
    \end{subfigure}
    \hfill
    \begin{subfigure}[t]{0.49\textwidth}
	\includegraphics{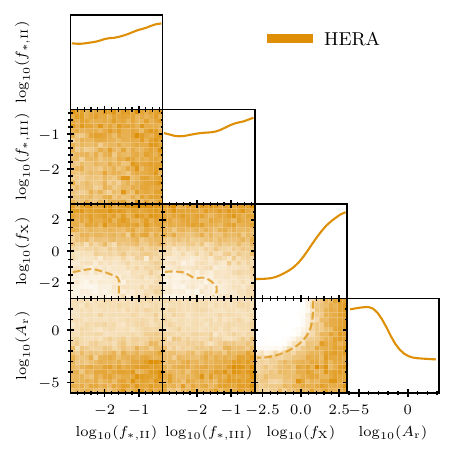}
        \phantomsubcaption
    \end{subfigure}
    \begin{subfigure}[t]{0.49\textwidth}
	\includegraphics{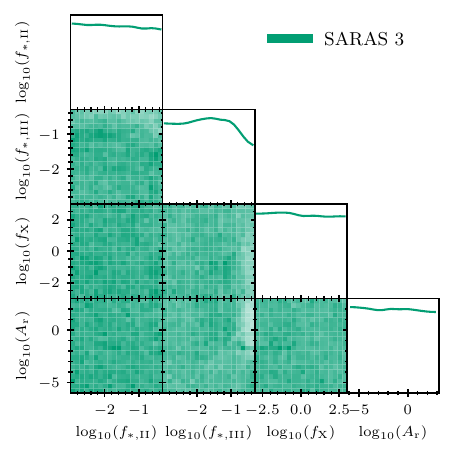}
        \phantomsubcaption
    \end{subfigure}
    \hfill
    \begin{subfigure}[t]{0.49\textwidth}
	\includegraphics{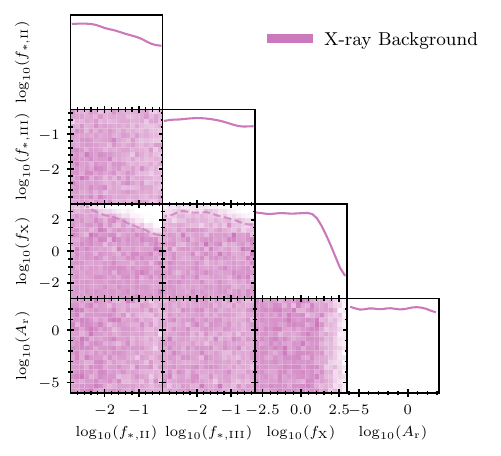}
        \phantomsubcaption
    \end{subfigure}
    \caption{Constraints on key parameters from our joint analysis (top left, blue), HERA (top right, orange), SARAS~3 (bottom left, green), and the unresolved X-ray background (bottom right, pink). 
    Shown are the 1D and 2D posteriors for the $\fstarII$, $\fstarIII$, $\fX$, and $\ar$ parameters. 
    On selected 2D posteriors 95\% confidence contours are shown as dashed lines to highlight disfavored or ruled-out regions. 
    Each of the three individual experiments rules out different regions of parameter space. 
    HERA rules out a corner of low $\fX \lesssim 1$ and high $\ar \gtrsim 0.01$ values, SARAS~3 rules out the combination of high $\fstarIII \gtrsim 0.15$ and high $\ar \gtrsim 0.1$, and the X-ray background strongly disfavouring high $\fX \gtrsim 100$ values, in particular in conjunction with high $\fstarII$. 
    The joint analysis rules out a significantly larger portion of the total parameter space than any of the individual constraints showing their complementarity as probes of the early Universe.
    This and subsequent figure were produced using \textsc{anesthetic}~\citep{anesthetic}.} \label{fig:individual_experiments}
\end{figure*}

The tightest constraints we see in our analysis are on combinations of the parameters $\ar$, $\fstarII$, $\fstarIII$, and $\fX$.
This is similar to what was found in previous studies~\citep{Bevins_2022, HERA_theory_22} that investigated models with synchrotron excess radio backgrounds, though those studies had a single star formation efficiency parameter. 
We depict these constraints for the individual experiments in Fig.~\ref{fig:individual_experiments}. 
Such a result was expected due to these parameters having the greatest impact on the 21-cm signal and X-ray background. 
Higher $\ar$ enhances the 21-cm signal magnitude as was seen in Fig.~\ref{fig:arad_variation_21cm}, while strong X-ray heating caused by high values of $\fstarII \fX$ or $\fstarIII \fX$ typically suppresses the magnitude of the 21-cm signal.
Furthermore, the interplay of $\fstarII$, $\fstarIII$, and $\fX$ determines both the magnitude of the X-ray background and the timing and strength of key features in the 21-cm signal, for example, the power spectrum heating peak.

Considering each experiment in turn we find, the HERA upper limits act to rule out a corner of models in $\ar$\,$-$\,$\fX$ space with high $\ar$ values $\gtrsim 0.01$ and low $\fX$ values $\lesssim 1$.
This region corresponds to strong radio backgrounds and weak heating, hence large 21-cm power spectra values, thus leading to this region being excluded by the HERA 21-cm power spectrum limits.
In addition, HERA disfavours but does not rule out, the combination of either low $\fstarII$ and low $\fX$, or low $\fstarIII$ and low $\fX$, since these also correspond to weaker heating.
The combinations of either high $\emin$ and high $\ar$, or low $\fX$ and high $\vc$, are also seen to be disfavored, discussed further in appendix~\ref{app:astro}, and are similarly attributable to weak X-ray heating, strong radio backgrounds, or both.  
From the marginalized $\ar$ posterior we find HERA disfavours $\ar$ values above $0.03$ at 68\% confidence.

\begin{figure}
    \centering
    \includegraphics[width=0.48\textwidth]{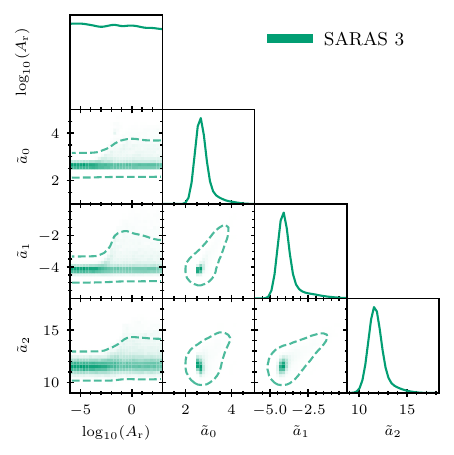}
    \caption{Posterior distribution of $\ar$ and the first three SARAS foreground parameters from the SARAS~3 data analysis.
For plotting clarity, the foreground parameters have been normalized via \mbox{$\Tilde{a}_{\rm 0} = 10^4 a_{\rm 0} - 35440$}, \mbox{$\Tilde{a}_{\rm 1} = 10^4 a_{\rm 1} + 2190$}, and \mbox{$\Tilde{a}_{\rm 2} = 10^4 a_{\rm 2}$}. 
As these transformations are linear, the priors over the transformed parameters remain uniform, and the correlations with other parameters remain unaffected. 
The 2D posteriors between $\ar$ and the foreground parameters show correlations for $\ar$ values in the range $0.001$ to $10$.
These correlations are most evident in the top 95\% confidence contour (dashed line) in each panel, along which it can be seen that the foreground parameters increase to compensate for the deeper 21-cm signals that can be produced at high $\ar$ values.
As expected, since high $\ar$ does not always produce deep 21-cm global signals, the lower 95\% confidence contours in these panels only show a weak change with $\ar$.
The lack of correlations outside the above $\ar$ range is due to the 21-cm signal not being significantly enhanced for $\ar$ values below this range, and the impacts of high $\ar$ saturating above this range as seen in Fig.~\ref{fig:arad_variation_21cm} and discussed in Section~\ref{sec:21cm}.
    }
    \label{fig:foreground_corr}
\end{figure}

The SARAS~3 results are less constraining, with the principle constraint from SARAS~3 being a ruling out of a small region of high $\ar \gtrsim 0.1$ and high $\fstarIII \gtrsim 0.15$  parameter space.
This region corresponds to vigorous early star formation with strong radio backgrounds.
SARAS~3 rules out such models as they predict a deep absorption trough minimum within its redshift range of 15.7 to 24.8.
The lack of constraints from SARAS~3 on the $\ar$\,$-$\,$\fX$ parameter space is somewhat surprising given the constraints \citet{Bevins_2022} found on a similar synchrotron excess radio background model using the same data. 
We find the principal difference between the two models that leads to the weaker constraining power of the SARAS~3 measurements is the change of radio background exponents, $\br = 2.6$ in their study, to $\br = 2.25$ in ours.
The weaker evolution of the excess radio background in our study leads to smoother and later 21-cm global signal minima. 
As a result, many high $\ar$ low $\fX$ models are not ruled out as might be expected due to the global signal absorption minimum moving outside the SARAS~3 band (as was shown in Fig.~\ref{fig:arad_variation_21cm}), making the
global signal smooth across the SARAS~3 band and thus indistinguishable from the galactic foregrounds.
Consequently, in these cases, the deep 21-cm signal can also be fit by the foreground model leading to a degeneracy between foreground parameters and 21-cm signal parameters.
This is confirmed via our finding of correlations in our 2D posteriors between $\ar$ and the foreground parameters $a_{\rm 0}$, $a_{\rm 1}$, and $a_{\rm 2}$, shown in Fig.~\ref{fig:foreground_corr}, whereas \citet{Bevins_2022} found their astrophysical and foreground parameter spaces to be independent. 
Due to this degeneracy, both the astrophysical parameter and foreground parameter constraints we find from SARAS~3 are weakened.

As is anticipated, the constraints from the X-ray background measurements are principally on the X-ray emission efficiency of (both Pop~II and Pop~III) star-forming halos $\fX$, with the highest $\fX$ values $>100$ almost entirely ruled out. 
The $\fstarII$\,$-$\,$\fX$ joint distribution shows the multiplicative degeneracy between these two parameters, with lower $\fstarII$ values ruled out as $\fX$ is increased. 
A weaker version of this trend is seen in the $\fstarIII$\,$-$\,$\fX$ joint posterior.
As there are far fewer total Pop~III stars than Pop~II stars the X-ray background rules out a smaller region in $\fstarIII$\,$-$\,$\fX$ than $\fstarII$\,$-$\,$\fX$ space.

We thus find that each of the individual observables rules out different regions of parameter space. 
This is reflected in our joint analysis results, where the complementarity of these experiments is made clear by a larger region of the total parameter space being ruled out or disfavored when compared to any of the experiments individually.
To quantify this statement more concretely, in our joint analysis, 51\% of the astrophysical prior is ruled out, whereas HERA, SARAS~3, and X-ray background measurements individually rule out 39\%, 6\%, and 5\%, respectively (see appendix~\ref{ssec:quantifying_constraints} for mathematical details). 
Most importantly for this study X-ray background measurements disfavouring high $\fX$ when combined with the HERA constraints on $\ar$\,$-$\,$\fX$ rule out a portion of high $\ar$ space that was allowed by HERA alone.
Consequently, the disfavouring of high $\ar$ above $\sim 0.01$ is enhanced through the joint analysis.

\begin{figure}
    \centering
    \includegraphics{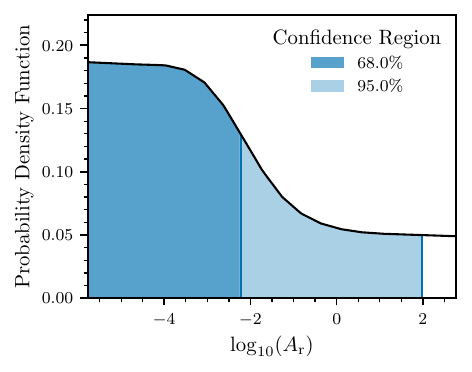}
    \caption{Joint analysis constraints on $\ar$, in the form of the posterior on $\log_{\rm 10}(\ar)$ (black line). 
    High $\ar$ $\gtrsim 1$ are seen to be disfavored compared to low $\ar$ $\lesssim 0.001$, however, the posterior never falls close to zero, and so no regions of $\ar$ can be confidently ruled out. 
    The 68\% and 95\% upper confidence regions are illustrated in dark blue and light blue showing the favouring of $\ar$ values $< 0.006$ and $< 95$ respectively. 
    Since no values of $\ar$ can be confidently ruled out we conclude current 21-cm signal data is insufficient to rule out any superconducting cosmic string models.}
    \label{fig:arad_cis}
\end{figure}

Since the focus of this study is on superconducting cosmic strings and thus $\ar$, from here on in, we concentrate on our $\ar$ constraints and their implications rather than on the astrophysical parameters.
For a detailed discussion on the insights that these data sets (and others) provide on high redshift astrophysics, see \citet{Pochinda_2023} (appendix~\ref{ssec:astro_parameter_constraints}, also provides a brief discussion on our astrophysical constraints). 
The marginalized posterior on $\log_{\rm 10}(\ar)$ from our joint analysis is shown in Fig.~\ref{fig:arad_cis}.
Values of $\ar$ below $\sim 0.001$ are found to be fully consistent with all three considered data sets.
Higher $\ar$ values above this threshold become increasingly disfavoured, the posterior gradually decreasing,  until it plateaus around $\ar \sim 1.0$.
We find these high $\ar \gtrsim 1.0$ values to be less than a third as likely as the low $\ar \lesssim 0.001$ values in light of the data. 
However, notably our $\log_{\rm 10}(\ar)$ posterior is not close to zero in any region, and so no $\ar$ values are confidently ruled out.

We hence can conclude that no strong constraints can be established on superconducting cosmic strings from current 21-cm signal data, including from the latest HERA upper limits. 
This is in contrast to the findings of \citet{Brandenberger_2019} who inferred constraints on the superconducting cosmic string parameters from the EDGES signal. 
The principle difference in our analysis is our inclusion of astrophysical uncertainties, in particular the unknown X-ray emission efficiency $\fX$ of early Universe galaxies.
As can be seen in Fig.~\ref{fig:individual_experiments} there remains a region of high $\fX$ and high $\ar$ values for which X-ray heating suppresses the 21-cm signal enhancement effect from the excess radio background, making these signals consistent with current 21-cm signal measurements. 
Thus the presence of uncertain astrophysical heating prevents constraints on cosmic strings from HERA or SARAS~3 data. 

\begin{figure}
    \centering
    \includegraphics{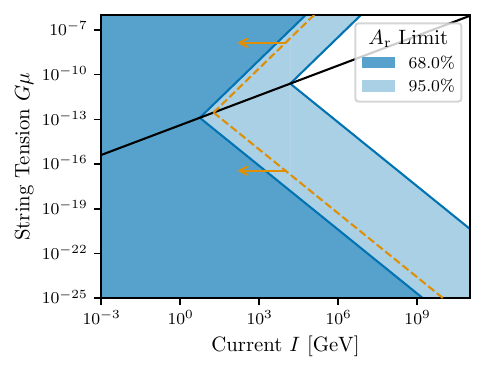}
    \caption{$\ar$ confidence regions translated to superconducting cosmic strings parameter space. Here we assume $\beta = 10$, $\gamma = 10$, $\kappa = 1$, $\Tilde{\kappa} = 1$, and $\chi = 10$~\citep{Theriault_2021}. The dark (light) blue region has an $\ar$ value less than the 68\% (95\%) upper confidence values from Fig.~\ref{fig:arad_cis} of $0.006$ ($95$). Also shown as an orange dashed line are the constraints from \citet{Brandenberger_2019} assuming $\mathcal{R} = 1$, orange arrows indicate the direction of the permitted region. We see the 68\% confidence $\ar$ contour corresponds roughly to the constraints previously found in \citet{Brandenberger_2019}. The 95\% confidence $\ar$ contour shows even if we were to treat our disfavored region of $\ar$ as ruled out, which we do not advocate, the constraints on superconducting cosmic strings from current 21-cm signal data would be orders of magnitude weaker than have been previously stated.}
    \label{fig:final_scs_constraints}
\end{figure}

There are, however, regions of $\ar$, and thus classes of superconducting cosmic strings, that are disfavoured by our analysis. 
We find upper confidence regions on $\ar$ of $< 0.006$ at the 68\% level and $< 95$ at the 95\% level,\footnote{Due to the ongoing debate about the existence of CMB heating~\citep{Venumadhav_2018, Meiksin_2021}, we performed a separate analysis with CMB heating disabled. The resulting conclusions did not change, and the confidence region values remained the same to within a few per cent. } also illustrated in Fig.~\ref{fig:arad_cis}. 
By assuming values for the dimensionless cosmic string model-specific parameters $\beta$, $\gamma$, $\kappa$, $\Tilde{\kappa}$, and $\chi$, we can translate these values to regions in the superconducting cosmic string parameters space of tension $\mu$ versus current $I$. 
Using the values from \citet{Theriault_2021}, $\beta = 10$, $\gamma = 10$, $\kappa = 1$, $\Tilde{\kappa} = 1$, and $\chi = 10$, which were motivated by theoretical expectations and numeric simulations, we find the favoured and disfavoured regions illustrated in Fig.~\ref{fig:final_scs_constraints}.
Also shown on this figure in an orange-dashed line are the $\mathcal{R} = 1$ constraints from \citet{Brandenberger_2019}.

The region we find to be disfavored based on our 68\% confidence value ($\ar > 0.006$) corresponds roughly to that \citet{Brandenberger_2019} argued was excluded.
Taken together this would seem to be indicative that this is the level of constraints that could potentially be achieved from future 21-cm signal measurements.
At the 95\% level, the disfavored region corresponding to $\ar > 95$ is several orders of magnitude higher in string current than the previously estimated constraints.
To reiterate neither of these should be taken as constraints on superconducting cosmic strings as we do not convincingly rule out any of the $\ar$ values we consider, but are just given as a comparison to previous works and to facilitate speculation about the prospects for future constraints. 

Due to the fact we are already observing a disfavouring of high $\ar$ values the prospects for constraints on superconducting cosmic strings from 21-cm signal observations soon seem promising.
We found the HERA Phase 1 21-cm power spectrum upper limits~\citep[full 94 night season,][]{HERA_obs_23} to be the most constraining part of our joint analysis and HERA is still in operation. 
So improved upper limits on the 21-cm power spectrum or even a detection, are expected in the coming years from the later phases of their observations, which shall also cover lower frequencies (higher redshifts). 
In addition, proposed upcoming X-ray telescopes such as \textit{AXIS}~\citep{AXIS}, \textit{ATHENA}~\citep{ATHENA}, or \textit{LYNX}~\citep{LYNX}, may be able to resolve further point sources, reducing the unresolved X-ray background and thus the upper limits on high redshift X-ray background. 
Finally, ongoing broader band 21-cm global signal experiments such as REACH~\citep{REACH} and MIST~\citep{MIST}, should avoid the issues we encountered using the SARAS~3 data wherein the global signal minimum moves outside the band, and thus lead to more substantial constraints on $\ar$ from 21-cm global signal null detections, or provide a definitive global signal detection.

\section{Discussion and Conclusions}~\label{sec:conclusions}

In this paper, we have performed a joint Bayesian analysis on the latest 21-cm signal data and measurements of the unresolved X-ray background, in an attempt to constrain superconducting cosmic strings. 
Specifically, we consider, both individually and together, the constraints imposed by the SARAS~3 21-cm global signal experiment~\citep{SARAS3}, the HERA Phase 1 21-cm power spectrum upper limits~\citep{HERA_obs_23}, and combined X-ray background measurements from several satellites including \textit{CHANDRA}~\citep{Hickox_2006, Harrison_2016}. 
We believe our analysis constitutes the first attempt to use the 21-cm power spectrum to constrain cosmic strings. 
Ultimately, we conclude that no such constraints can currently be inferred from the 21-cm signal, primarily due to the potential for heating by X-ray sources masking the impacts of the strings on the 21-cm signal. 

We adopted the model of \citet{Brandenberger_2019} for the excess radio background from superconducting cosmic strings, integrating it into the semi-numerical 21-cm signal code \textsc{21cmSPACE}.
This allowed for the simulation of the 21-cm signal and X-ray background for different superconducting cosmic string parameters and astrophysical scenarios. 
Hence, facilitating constraints using the HERA, SARAS~3, and unresolved X-ray background measurements individually, and as a joint fit. 

Due to the uncertainties surrounding the astrophysical parameters of the first galaxies, we followed a nested sampling-based Bayesian approach for our analysis.
Such a method allowed us to rigorously marginalize the astrophysical and foreground parameters away as nuisance parameters.
Since \textsc{21cmSPACE} has a runtime of a few hours, we could not practically use it directly in our analyses.
We instead trained neural network emulators on the outputs of 44,836 simulations and utilized these for our analysis, representing a $10^{3.5}$ times reduction in the number of simulations required than if we had directly called \textsc{21cmSPACE} in our joint analysis, thus making our Bayesian approach feasible.

Analysis of the data sets individually revealed: the HERA 21-cm power spectrum upper limits rule out the combination of a strong radio background from cosmic strings and weak X-ray heating; the SARAS~3 21-cm global signal data eliminates a small area of strong radio backgrounds and efficient Pop III star formation; and X-ray background measurements rule out the highest rates of X-ray heating. 
By combining these data sets in our joint analysis we found a disfavouring of a strong excess radio background, and hence a region of high string current and high string tension parameter space.
However, while this region was disfavored at 68\% confidence, no part of the string parameter space was ruled out.
We thus conclude the latest 21-cm cosmology data, at the moment of writing, gives no constraints on cosmic strings.

The difference between our conclusions and that of previous studies~\citep[e.g.][]{Brandenberger_2019} principally stem from our inclusion of the astrophysical uncertainties. 
Primary amongst these is the uncertainty in X-ray heating from high redshift sources, as efficient X-ray heating acts to suppress the 21-cm signal and thus mask any impacts of cosmic strings.
If we knew for certain the X-ray emission efficiency of high redshift sources was  $\lesssim 3 \times 10^{40}$\,erg\,s\,yr$^{-1}$\,M$_{\odot}^{-1}$, then our analysis would predict similar cosmic string constraints to that of \citet{Brandenberger_2019}.
However, theoretical expectations are that high redshift X-ray sources have efficiencies of $3 \times 10^{40}$\,erg\,s\,yr$^{-1}$\,M$_{\odot}^{-1}$~\citep{Fragos_2013} or greater~\citep{Sartorio_2023, Liu_2023}, values for which we find no constraints on the strength of the radio background and hence no constraints on cosmic strings. 
This is of course all assuming the excess radio background is dominated by superconducting cosmic strings as we have done throughout this paper.
The inclusion of other sources of excess radio background, such as high redshift radio galaxies~\citep{Reis_2020}, would weaken any constraints on cosmic strings and thus further strengthen our conclusions.
We provide additional discussion on the astrophysical constraints from this work in appendix~\ref{app:astro}.

As we find some regions of cosmic string parameter space are already disfavoured the prospects for near-future constraints on superconducting cosmic strings from 21-cm cosmology seem promising.
Ongoing or upcoming experiments that could provide better limits or definitive detections of the 21-cm power spectrum include HERA~\citep{HERA_obs_23}, NenuFAR~\citep{NenuFAR}, LOFAR~\citep{LOFAR}, and SKA1-LOW~\citep{Koopmans_2015}, or for the 21-cm global signal, REACH~\citep{REACH}, PRIZM~\citep{PRIZM}, MIST~\citep{MIST}, SARAS~4, and EDGES~3. 
The broader bands of some of these upcoming global signal experiments should avoid the issue we encountered when attempting to extract cosmic string constraints from SARAS~3 data, that the global signal minima moved outside of the science band.
In addition, the overdensity produced by cosmic string wakes may be able to be directly observed in SKA 21-cm signal images or detected in 21-cm three-point statistics~\citep{Brandenberger_2010, Hernandez_2012, Hernandez_2014, Maibach_2021}.
Looking to the 2030s and beyond there is also the potential for improved X-ray background measurements from proposed next-generation X-ray observatories, \textit{AXIS}~\citep{AXIS}, \textit{ATHENA}~\citep{ATHENA}, or \textit{LYNX}~\citep{LYNX}, that could be used in a future joint analysis.
Any constraints from 21-cm cosmology on cosmic strings could then be used to supplement those from spectral distortions~\citep{Tashiro_2012}, pulsar timing arrays~\citep{Quelquejay_2023}, big bang nucleosynthesis~\citep{Miyamoto_2013}, reionization history, or the unexplained radio background~\citep{Cyr_2023a, Cyr_2023b} in a joint analysis, or provide an independent probe to confirm a discovery.

While this work was in its final stages of preparation, papers by \citet{Acharya_2023} and \citet{Cyr_2023a} were announced, proposing a new mechanism by which superconducting cosmic strings can impact the 21-cm signal, \textit{soft photon heating}. 
These studies find that this heating from low frequency ($\lesssim 12$ MHz at $z = 20$) radio photons diminishes the magnitude of the 21-cm global signal absorption trough and thus prevents them from extracting any constraints on cosmic strings when treating the claimed EDGES detection of the global 21-cm signal as a limit on the absorption trough depth. 
As we have found that astrophysical heating is already sufficient to prevent any meaningful constraints being placed on superconducting cosmic strings from current 21-cm signal data, including this additional heating would have strengthened our conclusions.

Should a definitive 21-cm signal detection be made, soft photon heating may prove advantageous for extracting constraints on superconducting cosmic strings. 
A cosmic string produced excess radio background on its own would be challenging to distinguish from cooling from interacting dark matter or a radio background from other sources.
However, the combination of a radio background and soft photon heating both from cosmic strings and hence with linked time evolutions, should break many degeneracies allowing for firmer conclusions to be drawn. 
In addition, soft photon heating being effectively uniform is in stark contrast to astrophysical sources of heating which are anticipated to be clustered around overdensities. 
The distribution of heating of the early Universe is inferable from the variation of the 21-cm power spectrum across wavenumber~\citep{Gessey-Jones_2023}, hence providing another signature to search for cosmic strings in the 21-cm signal.
Thus while in this study we find no constraints on superconducting cosmic strings from 21-cm data, the outlook for doing so in the future appears optimistic.

\section*{Acknowledgements}

The authors are grateful to Stefan Heimersheim and Irene Abril-Cabezas for sharing their power spectrum emulator and HERA data analysis codes.
We also wish to thank Petra Brcic, Roxane Thériault, Jordan Mirocha, and Robert Brandenberger for helpful conversations about the superconducting cosmic string excess radio background and the anonymous referee whose comments greatly aided in improving this paper.

TGJ acknowledges the support of the Science and Technology Facilities Council (UK) through grant ST/V506606/1. 
SP is grateful to the Cambridge Trust and the Centre for Doctoral Training in Data Intensive Science for their support through a Cambridge International studentship. 
HTJB acknowledges support from the Kavli Institute for Cosmology Cambridge, the Kavli Foundation, and the Science and Technology Facilities Council (UK) through grant ST/T505997/1. 
AF and WJH thank the Royal Society for their support through their University Research Fellowships. 
EdLA is grateful to the Science and Technology Facilities Council (UK) for their continued support through his Rutherford Fellowship.
RB is thankful for the support of the Israel Science Foundation (grant No.\ 2359/20).

This work used the DiRAC Data Intensive service (CSD3, project number ACSP289) at the University of Cambridge, managed by the University of Cambridge University Information Services on behalf of the STFC DiRAC HPC Facility (\href{www.dirac.ac.uk}{www.dirac.ac.uk}). The DiRAC component of CSD3 at Cambridge was funded by BEIS, UKRI and STFC capital funding and STFC operations grants. DiRAC is part of the UKRI Digital Research Infrastructure.

For the purpose of open access, the author has applied a Creative Commons Attribution (CC BY) licence to any Author Accepted Manuscript version arising from this submission.

\section*{Data Availability}

The nested sampling inference products (e.g.\ chains) computed in this work are made available at \href{https://zenodo.org/record/8362801}{https://zenodo.org/record/8362801} alongside code to reproduce most of the figures presented from said inference products. All other data and codes produced in this work are available on reasonable request to the corresponding author.



\bibliographystyle{mnras}
\bibliography{scs_paper} 



\appendix

\section{Astrophysical Constraints} \label{app:astro}

While the concentration of this study was on superconducting cosmic strings, our nested sampling-based methodology also produces astrophysical constraints as a by-product. 
These were briefly discussed in Section~\ref{sec:results}.
Here we go into more detail about the constraints on each astrophysical parameter (subsection~\ref{ssec:astro_parameter_constraints}).
In addition, we present the functional posteriors on each of the observables employed (subsection~\ref{ssec:functional_posteriors}) and quantify which data set is the most constraining in our analysis (subsection~\ref{ssec:quantifying_constraints}).

\subsection{Astrophysical parameter constraints}~\label{ssec:astro_parameter_constraints}

\begin{figure*}
    \centering
    \includegraphics[width=\textwidth]{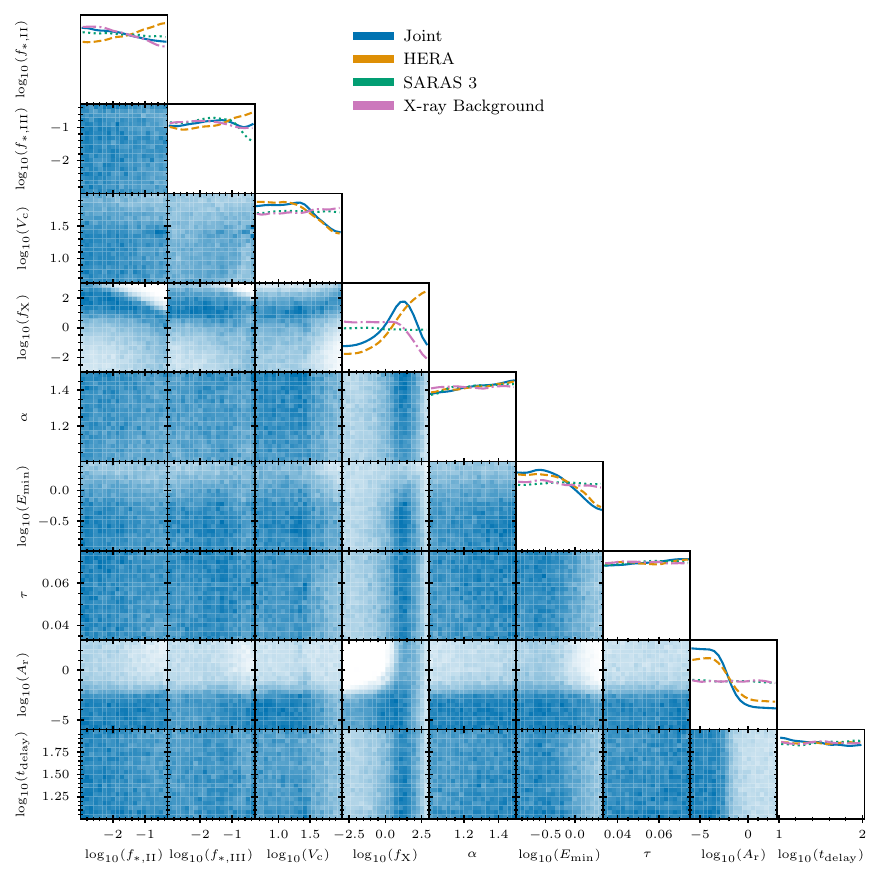}
    \caption{Constraints from our joint analysis on the astrophysical parameters and $\ar$. A subset of this corner plot was depicted in Fig.~\ref{fig:individual_experiments}. The 2D posteriors depict only the constraints from the joint analysis, whereas the 1D posteriors show the joint analysis (blue solid) alongside the results from HERA (orange dashed), SARAS~3 (green dotted) and the X-ray background (pink dot-dashed). In the 1D posteriors moderate constraints are seen on $\fX$ and $\ar$, with weaker constraints on $\fstarII$, $\vc$ and $\emin$. No constraints are seen on $\tau$, $\alpha$ or $\tdel$. We find that the combination of $\emin \gtrsim 1.0$\,keV and $\ar \gtrsim 0.1$ is ruled out due to high $\emin$ leading to inefficient X-ray heating and thus strong 21-cm signals. Since both $\fX$ and $\emin$ impact the efficiency of X-ray heating a degeneracy between them is apparent in the disfavored region of their joint posterior. Similarly, due to the combination leading to weak X-ray heating, we also observe the combination of $\vc \gtrsim 30$ and $\fX \lesssim 1$ to be disfavoured.}
    \label{fig:combined_corner_plot}
\end{figure*}

The constraints on all astrophysical parameters and $\ar$ from our joint analysis are shown in Fig.~\ref{fig:combined_corner_plot}.
For ease of comparison on the 1D posteriors, we also depict the constraints from HERA, SARAS~3, and the X-ray background individually. 
Considering the 1D posterior and 2D posteriors of each astrophysical parameter we find:
\begin{itemize}
    \item $\fstarII$: A slight disfavouring of high $\fstarII$ is seen, principally driven by the X-ray background constraints. The multiplicative degeneracy between $\fstarII$ and $\fX$ is evident in their joint posterior, with $\fstarII \fX \gtrsim 5$ ruled out by the X-ray background, and $\fstarII \fX \lesssim 0.001$ disfavored by HERA. 
    \item $\fstarIII$: The 1D posterior of $\fstarIII$ shows no clear disfavored values due to the competing influences of the three observables largely cancelling. SARAS~3 can rule out deep early global signal minima leading to the disfavouring of the combination of $\fstarIII \gtrsim 0.15$ and $\ar \gtrsim 0.1$. The X-ray background rules out $\fstarIII \fX \gtrsim 50$, while HERA disfavours the strong heating from $\fstarIII \fX \lesssim 0.001$.
    \item $\vc$: Values of $\vc > 30$ are found to be slightly disfavoured due to HERA. The joint posteriors reveal this is driven by HERA disfavouring the combination of $\vc \gtrsim 30$ and $\fX \lesssim 1$. Since high $\vc$ values lead to lower star formation rates, this region would correspond to weak X-ray heating, which we have previously seen HERA rules out in combination with strong radio backgrounds, thus explaining these trends. 
    \item $\fX$: Both HERA and the X-ray background provide constraints on $\fX$. This can be seen clearly in the 1D posterior where HERA disfavours $\fX \lesssim 1$ while the X-ray background disfavours $\fX \gtrsim 100$, leading to the joint constraint moderately favouring $\fX \sim 10$. The joint posterior of $\fX$ and $\ar$ is the strongest constrained of the 2D joint posteriors in the astrophysical part of the parameter space, with a ruling out of $\fX \lesssim 1$ and $\ar \gtrsim 0.01$. In addition to the previously discussed constraints, we see a disfavouring of a region of the $\emin$\,$-$\,$\fX$ joint distribution where $\fX \sim 10$ and $\emin \gtrsim 1.5$\,keV. This constraint originates from HERA ruling out high radio backgrounds and weak X-ray heating. Higher $\emin$ leads to a harder X-ray spectrum and, since higher energy X-rays have a lower cross-section in the IGM, weaker X-ray heating.
    This also explains the apparent degeneracy between $\fX$ and $\emin$ on the low $\fX$ boundary of this disfavoured region as both high $\fX$ and low $\emin$ lead to increased X-ray heating. 
    \item $\alpha$: No constraints are seen on $\alpha$.
    \item $\emin$: Due to HERA disfavouring weak X-ray heating and strong radio backgrounds, we see a disfavouring of high $\emin$ values, and the ruling out of the combination of $\ar \gtrsim 0.1$ and $\emin \gtrsim 1.0$\,keV. For reference, early Universe X-ray binary spectra are predicted to have an $\emin$ in the range $0.1$\,keV to $1.0$\,keV~\citep{Fragos_2013, Sartorio_2023}.
    \item $\tau$: No constraints are seen on $\tau$. This is consistent with previous studies that constrained early Universe astrophysics with HERA Phase 1 upper limits~\citep[e.g.][]{HERA_theory_22, HERA_obs_23, Bevins_2023}.
    \item $\tdel$: No constraints are seen on $\tdel$.
\end{itemize}
\citet{Pochinda_2023} previously found constraints on $\fstarII$, $\fstarIII$, $\fX$, and $\vc $ from the same data sets considered here, however, no constraints on $\emin$ were reported in that study.
Our discovery of constraints on $\emin$ suggests future analyses may want to adopt the methodology utilized here, and in \citet{Bevins_2022}, of allowing neural network emulators to interpolate between discrete values and then sampling over the parameter continuously to gain additional insights into the astrophysics of the early Universe.

\subsection{Functional posteriors of observables}~\label{ssec:functional_posteriors}

An alternative perspective on our constraints is to view the functional posteriors of the observables themselves\footnote{Computed and depicted using our emulators and \textsc{fgixenx}~\citep{fgivenx}.} rather than the parameter posteriors. 
These allow us to verify that our posteriors are consistent with observations, as well as determine which data sets are the most constraining, and reveal where the most uncertainty remains in these observables, potentially informing future observations.

\begin{figure*}[htbp]
    \centering
    \includegraphics[width=\textwidth]{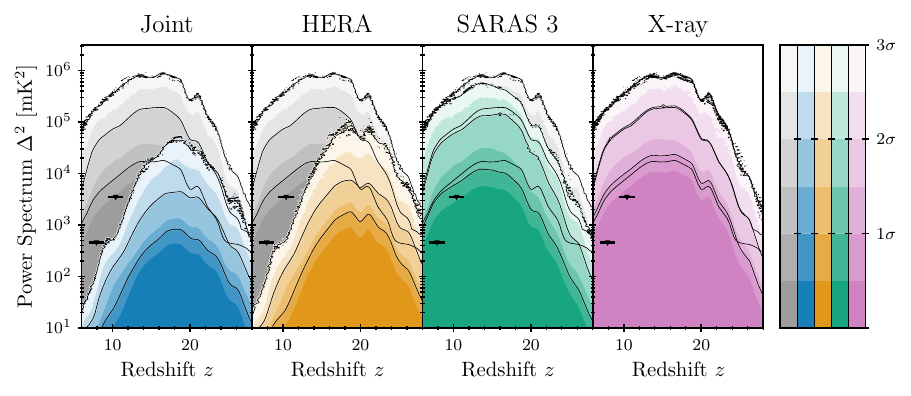}
    \caption{Functional posteriors on the 21-cm power spectrum at $k = 0.34$\,h\,cMpc$^{-1}$ ($0.23$\,cMpc$^{-1}$). The prior distribution is shown in grey, joint constraints posterior in blue, HERA posterior in orange, SARAS~3 posterior in green, and the X-ray background posterior in pink. No constraints on the 21-cm power spectrum are seen from the X-ray background, and the SARAS~3 constraints are weak, ruling out the strongest 21-cm power spectrum values at $z > 15$. The HERA data set constraints are strongest at lower redshifts $< 12$, with little constraining power at $z > 22$. Despite the SARAS~3 and X-ray background constraints being weak the joint analysis is not completely dominated by HERA, with the other two experiments improving the power spectrum constraints in the aforementioned high redshift region. The headline 21-cm power spectrum upper limits from \citet{HERA_obs_23} of $\Delta^2(k = 0.34\textrm{\,h\,Mpc}^{-1}) \leq 457$\,mK$^2$ at z = 7.9 and $\Delta^2(k = 0.36\textrm{\,h\,Mpc}^{-1}) \leq 3496$\,mK$^2$ at z = 10.4, are shown on the figure as black markers. The joint and HERA constraints posteriors are seen to be consistent with these limits, providing a sanity check for our methodology.  }
    \label{fig:power_functional_posterior}
\end{figure*}

The functional posterior of the 21-cm power spectrum at \mbox{$k = 0.34$\,h\,cMpc$^{-1}$} ($0.23$\,cMpc$^{-1}$) is depicted in Fig.~\ref{fig:power_functional_posterior} for each of our four analyses.
As was expected, we find the HERA data to be the most constraining on the 21-cm power spectrum and thus to dominate the joint constraints. 
SARAS~3 rules out strong 21-cm power spectra at redshifts $z > 15$ since these would be accompanied by a deep global signal in the SARAS~3 band. 
The X-ray background provides no constraint on the power spectrum.
Overall in the joint constraint, we find a large contraction of the allowed 21-cm power spectrum space from the prior to posterior showing the statistical power of the existing HERA upper limits.
A comparison of the posterior and prior also reveals the greatest uncertainty remains in the 21-cm power spectrum at high redshifts, motivating 21-cm power spectrum observations targetting $z > 15$ such as those underway by NenuFAR~\citep{NenuFAR} and OVRO-LWA~\citep{OVRO-LWA}.

\begin{figure*}
    \centering
    \includegraphics[width=\textwidth]{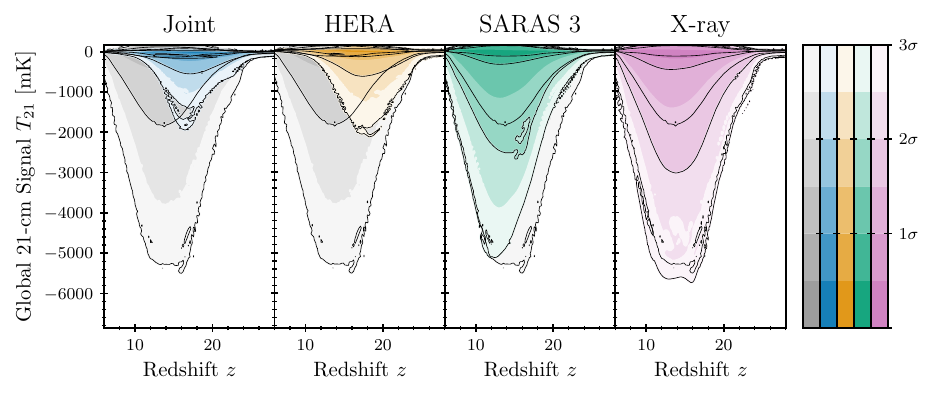}
    \caption{Functional posteriors on the 21-cm global signal. The prior distribution is shown in grey, joint constraints posterior in blue, HERA posterior in orange, SARAS~3 posterior in green, and the X-ray background posterior in pink. As was seen in Fig.~\ref{fig:power_functional_posterior} for the 21-cm power spectrum, no constraints are seen on the 21-cm global signal from the X-ray background. SARAS~3 provides moderate constraints on the 21-cm global signal, strongest in the SARAS~3 science band of redshift 15.7 to 24.8. The HERA power spectrum measurements are found to be quite constraining on the 21-cm global signal below $z = 20$ due to the link between the power spectrum magnitude and the global 21-cm signal. Our joint analysis shows a large contraction between posterior and prior, ruling out global signal amplitudes deeper than $-600$\,mK at 2$\sigma$ and $-2000$\,mK at 3$\sigma$. }
    \label{fig:global_functional_posterior}
\end{figure*}

In Fig.~\ref{fig:global_functional_posterior} we show the equivalent 21-cm global signal functional posteriors. 
Again the X-ray background provides minimal constraints.
The constraints from SARAS~3 are weak, ruling out only the deepest 21-cm signals at redshift 12 to 24, with no contraction between prior and posterior for $z < 12$.
HERA provides the strongest 21-cm global signal prior to posterior contraction, ruling out the deepest global signals up to $z \sim 20$, and 
strongly constraining the global signal below the HERA observation band ($z < 11.1$).
The  $z < 20$ constraints from HERA and $z > 12$ constraints from SARAS~3 lead to a large contraction between the prior and posterior in the joint analysis.  
The deepest permitted global signals from said joint analysis have their minimum between $z = 14$ to $18$, suggesting future global signal experiments should aim to cover this redshift range.

\begin{figure*}
    \centering
    \includegraphics[width=\textwidth]{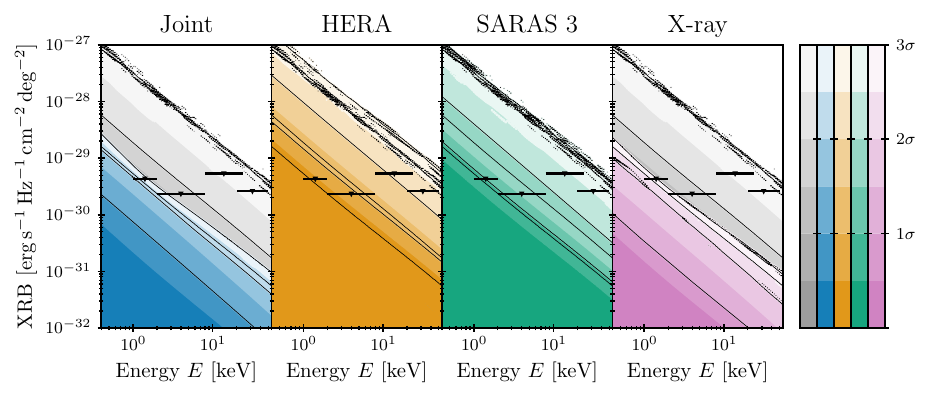}
    \caption{Functional posteriors on the X-ray background (XRB) from high redshift sources. The prior distribution is shown in grey, joint constraints posterior in blue, HERA posterior in orange, SARAS~3 posterior in green, and the X-ray background posterior in pink.  X-ray background constraints are only seen to come from direct X-ray background measurements, with the HERA and SARAS~3 observations providing no constraints. The integrated unresolved X-ray background values listed in Table~\ref{tab:ixrb_bands} are shown on the plot as black triangle markers, with the corresponding energy band indicated by a black bar. We find as expected that the X-ray background constraint and joint constraint posteriors are consistent with these upper limits whereas the HERA and SARAS~3 posteriors are not.}
    \label{fig:xray_background_posterior}
\end{figure*}

Finally, we find HERA and SARAS~3 have no significant constraining power on the X-ray background, with their functional posteriors nearly identical to the prior. 
As a result, the joint analysis constraints are entirely dominated by the X-ray background measurements, see Fig.~\ref{fig:xray_background_posterior}.

\subsection{Quantifying constraining power}~\label{ssec:quantifying_constraints}

Previously, we have found qualitatively that of the data sets we consider, the HERA power spectrum upper limits provide the strongest constraints on high redshift astrophysics and $\ar$.
However, a joint analysis still leads to a substantial improvement in constraining power over individual experiment constraints. 
Here, we quantify these statements more concretely. 

In the case of a top-hat prior and posterior, there is an intuitive geometric measure of the constraining power of a data set, the percentage of the parameter space volume that is consistent with the data ($\text{\%}_{\rm cons}$) (i.e.\ the parameter space volume of the posterior over the parameter space volume of the prior).
Using $\text{\%}_{\rm cons}$, the constraining power of data sets can then be compared, with lower percentages indicating greater constraining power.

Unfortunately, if the posterior (or prior) is not a top-hat distribution, its parameter space volume is no longer well-defined as there is no unambiguous inside and outside of the distribution in parameter space. Consequently, the above definition of $\text{\%}_{\rm cons}$ breaks down. 
However, the notion of the percentage of the prior that is consistent with the data can still be defined using information theory.
The amount of information gained through an experiment or observation is given by the Kullback–Leibler divergence $\mathcal{D}_{\rm KL}$ between posterior and prior~\citep{Sivia_2006}.
Additionally, one bit of information is the information gained from ruling out half of the prior possibilities~\citep{Mackay_2003}.
Combining these results, we arrive at an information-theoretic definition of the percentage of the prior that is consistent with the data\footnote{In this study, we use the $\log_{\rm e}$ convention for $\mathcal{D}_{\rm KL}$ and hence it is in units of nats rather than bits. If the $\log_{\rm 2}$ convention is used instead, the $e$ in this equation should be replaced with $2$.}
\begin{equation}~\label{eqn:quantifying-power} 
    \text{\%}_{\rm cons}  \equiv 100 e^{-\mathcal{D}_{\rm KL}},
\end{equation}
which has previously been used for quantifying the constraining power of data sets in \citet{Bevins_2022}.
Since $\mathcal{D}_{\rm KL} \geq 0$ with equality only when the prior and posterior are identical, this definition has the intuitive properties that it is always between 0\% and 100\%, with 100\% corresponding to the data having no constraining power. 
Furthermore, it can be shown that in the case of top-hat prior and posterior, it is equivalent to the parameter-space volume-based definition of $\text{\%}_{\rm cons}$ discussed previously.
However, unlike the volume-based definition, the information-theoretic definition remains well-defined between any prior and posterior distribution~\citep{Mackay_2003}.
Thus, equation~\ref{eqn:quantifying-power} gives a versatile and well-motivated quantification to the constraining power of each of our data sets.

We are, however, not interested in the constraining power of the datasets on all the parameters, only the astrophysical and $\ar$ parameters.
To address this, in equation~\ref{eqn:quantifying-power} instead of the full $\mathcal{D}_{\rm KL}$, we use the marginal Kullback–Leibler divergence between the astrophysical prior and posterior, where nuisance (e.g. foreground) parameters have been integrated away~\citep{Bevins_2023}.
Accordingly, we derive the percentage of the astrophysical prior consistent with the data and thus get a direct quantification of the constraining power of each dataset on the parameters of interest. 
In practice, we calculate the marginal posteriors on the astrophysical and $\ar$ parameters with \textsc{margarine}~\citep{margarine1, margarine2, margarine3}.

\begin{figure}
    \centering
    \includegraphics[width=0.48\textwidth]{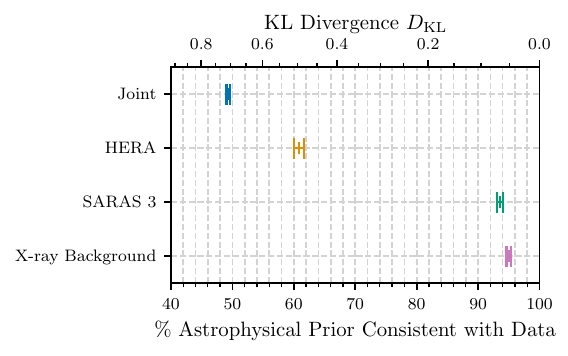}
    \caption{The percentage of the prior consistent with each data set and a joint analysis between them. The corresponding marginal Kullback–Leibler (KL) divergence is given on the top axis. The percentages of prior consistent with the data are found to be $49\%$ for the joint analysis (blue), $61\%$ for HERA (orange), $94\%$ for SARAS~3 (green) and $95\%$ for the X-ray background (pink). We hence see that HERA is the most constraining individual data set, but is not dominant, with a significantly larger prior contraction in the joint analysis than when using any of the experiments individually. }
    \label{fig:volume_contraction}
\end{figure}

The percentage of the astrophysical prior consistent with the data for each of the individual data sets is found to be $61\%$ for HERA, $94\%$ for SARAS~3 and $95\%$ for the X-ray background (see Fig.~\ref{fig:volume_contraction}). 
Lastly, for the joint analysis, the corresponding value is $49\%$.
This confirms our earlier qualitative assessment that HERA is the most constraining individual data set, ruling out $39\%$ of the  astrophysical prior whereas SARAS~3 and X-ray background are roughly equally constraining ruling out $\sim 5\%$ each. 
However, importantly they rule out different regions of parameter space and so a joint analysis of the three provides stronger constraints, with $51\%$ of the prior ruled out, retroactively justifying our joint analysis methodology.

We can also perform a similar analysis using the marginal Kullback–Leibler divergence on $\ar$ alone, to determine the extent to which each analysis constrains our parameter of interest. 
For such an analysis, we find that in our joint fit, $87\%$ of the $\ar$ prior is consistent with the data, with the equivalent values being $93\%$ for HERA, and $>99\%$ for SARAS~3 and the X-ray background. 
This is consistent with what would be expected based on our $\log_{\rm 10}(\ar)$ 1D posteriors and further illustrates the benefits of performing a joint analysis on these data sets.


\bsp	
\label{lastpage}
\end{document}